\documentclass[twocolumn,amssymb,aps,pra]{revtex4-1}

\newcommand{\etal}{\textit{et. al.}}

\usepackage{graphicx}
\usepackage{color,amsmath,amssymb}

\begin{document}

\title{Optimal control theory for rapid-adiabatic passage techniques in inhomogeneous external fields}%

\author{ Emil J. Zak$^{a}$ }
\email[Corresponding author: ]{emil.j.zak@gmail.com}
\affiliation{$^{a}$Department of Physics and Astronomy, University College London, Gower Street, WC1E 6BT, London}
\date{\today}%

\begin{abstract}
The present paper reports on results of quantum dynamics calculations for Stark-chirp rapid-adiabatic passage (SCRAP) in two-level systems with electric fields computed with the optimal control theory. The Pontryagin maximum principle is used to determine the robust optimal control fields in the presence of time-varying and spatially-inhomogeneous perturbing electric fields. The concept of a non-adiabatic correction to the Bloch vector dynamics is introduced and discussed. The existence of a non-zero geometric phase is proved for certain adiabatic paths, which correspond to the complete population return in the rapid-adiabatic passage. A connection is shown between the geometric phase and a measure of the non-adiabatic effects in the time evolution of the state vector during SCRAP. Different cost functionals used in the optimal control scheme are shown to correlate with different topologies of the paths followed by the parameters of the Hamiltonian, which tightly relates to the values of the geometric phase acquired by the adiabatic wavefunction.
\end{abstract}

\maketitle

\section{Introduction}
Rapid-adiabatic passage techniques are used in many areas of physics as a controllable and robust means of transforming quantum states. For their exceptional stability to minor uncertainties in the experimental parameters they gained recognition in the field of quantum control, predominantly finding applications in quantum computing \cite{Nie2010,Chen2015,10.2307/41739851,Shi2014,Dolde2014,Huang2014} and nuclear magnetic resonance \cite{Kobzar2008,Kobzar2012,VanReeth2017}.
One of the major challenges for the general class of quantum optimal control experiments (OCEs) \cite{Hornung2000,Ren2006,Nuernberger2007,Brif2011,Wells2013,Parker2014}, including rapid-adiabatic passage \cite{Strasfeld2007,Sugny2009,Schnfeldt2009,Dong2013,Brif2014}, are  perturbations to the system caused by the interactions with the environment, which can be both time-dependent and inhomogeneous in space. This problem is primarily relevant to manipulation of \textit{qubit} ensembles \cite{Nie2010,Chen2015,10.2307/41739851,Shi2014,Dolde2014,Huang2014}, but may also be important in alignment of molecules \cite{Pelzer2007,Suzuki2008,Artamonov2010}, high-accuracy spectroscopic measurements \cite{Chen2015,Lin2011} photo-induced chemical reactions \cite{Kurosaki2009,Krieger2011}, reactive scattering experiments \cite{Pan2017} or even magnetic resonance imaging \cite{Kobzar2012,VanReeth2017}.

Optimal control theory \cite{jurdjevic,Bonnard2012} has been successfully applied to various aspects of the rapid-adiabatic passage \cite{Werschnik2007,Reich2012,Riviello2017,Boscain2002,Assmat2012,VanDamme2017}. In particular, the Stark-chirp rapid-adiabatic passage (SCRAP) method \cite{Rickes2000,Yatsenko1999,Oberst2008,Shore2009} is of interest for realisation of dipole allowed transitions. 
The role of protocols for optimal control in SCRAP can be viewed as the maximization of the population transfer between the initial state of the system and the target state, subject to external perturbing electric fields. 

The present paper focuses on the problem of optimal control for SCRAP in the presence of time-varying and spatially inhomogeneous electric fields.
A two-level quantum system serves as a model for optimizing the process of Stark-chirped transformations of quantum states.  

The choice of rapid-adiabatic-passage techniques for the \textit{qubit} transformations is mainly dictated by their robustness and the low sensitivity to the total pulse energy \cite{Shore2009,Vitanov2017}, but most importantly because they can, in principle, provide 100\% population transfer between the initial and the target state \cite{Parker2012}. However, even robustness of SCRAP becomes limited when external perturbing electric fields are present; these fields can affect the Stark-chirp rate, which may lead to a decrease in the overall efficiency of the population transfer process.

Optimal control theory is known an excellent tool for predicting the optimal values for the experimentally tunable parameters. In particular, the Pontryagin Maximum Principle (PMP) \cite{jurdjevic,Bonnard2012} can be used in the search for optimal control fields \cite{Chen2011,Assmat2012,VanDamme2017}. In contrast to standard optimal control methods \cite{Werschnik2007,Reich2012,Riviello2017}, the PMP carries the advantage of finite-dimensional Hamilton-like dynamics, which minimizes a given cost functional, such as the energy of the pulse or the duration of the pulse. Such an approach is particularly suitable when the time-dependent Schrodinger equation for the \textit{qubit} is transformed into a set of first-order differential Bloch equations \cite{gerry_knight_2004}. Combining the dynamics on the Bloch sphere with the PMP gives a framework for control of the population transfer in the \textit{qubit}. Among a number of works aimed at unifying SCRAP with optimal control theory, utilization of the PMP is limited to few studies \cite{Boscain2002,Assmat2012,VanDamme2017}. Among those, Van Damme \etal\ \cite{VanDamme2017} considered perturbatively the problem of robustness of the optimal control fields to offset field inhomogeneities, which are constant in space and time.

From the practical perspective, there are two main reasons for which the efficiency of SCRAP can get lowered. First, a rapidly time-varying perturbation (inhomogeneity) can cause transitions between adiabatic states. Thus controlling the adiabaticity of time evolution of the state vector stands as one of the objectives for optimal control in SCRAP. The other situation is when a position-dependent external electric field induces static Stark-shifts in the system, which alter the Stark-chirp rate and produce a spread in the initial static detuning. Without optimization of the control fields, the global efficiency of SCRAP over a range of positions in space can be low. 
Although a number of studies was dedicated to optimization of the process of manipulation of spatially inhomogeneous ensembles \cite{Khaneja2005,Li2007,Altafini2007,Chen2014,Song2016}, no uses of the PMP have been reported so far, to the best of author's knowledge. 

For this reason, a general protocol for the optimal control in SCRAP is presented in this work. 
The relevance of the electric field inhomogeneities, which are time-dependent and spatially inhomogeneous is visible in various molecular motion control experiments \cite{Lazarou2010,Parker2011,Parker2012}, including Stark deceleration \cite{Bethlem2002PRL,Kupper2006} and Sisyphus cooling \cite{Zeppenfeld2012,Owens2017}. Wherever inhomogeneities in the external fields are present, an optimization scheme is necessary for achieving the complete population transfer. This paper is aimed at providing a ready-to-use methodology for these types of situations.

Section \ref{sec:theory} outlines the theoretical model used to simulate the quantum dynamics for the \textit{ qubit }in the SCRAP scheme. Next, in section \ref{sec:gauss}, the SCRAP dynamics is analysed and optimized in the adiabatic approximation with fixed Gaussian time profiles of pulses. Non-adiabatic effects are also discussed with a description of a method for quantifying the deviations from adiabatic dynamics. Section \ref{sec:general} presents an extension to the full dynamics calculations and an application of the PMP to space- and time- dependent perturbing fields. Both minimization of the total energy of the pulses and maximization of the adiabaticity of the process is considered. Finally the role of the geometric phase in rapid-adiabatic passage is discussed.

\section{Theoretical model}
\label{sec:theory}
\subsection*{A two-level quantum system}
The state vector for a two-level quantum system can be written as a linear combination of basis states:

\begin{equation}
|\Psi\rangle = c_1(t)|1\rangle + c_2(t)|2\rangle
\label{eq:wf}
\end{equation}
where $\mathbf{C}(t)=[c_1(t),c_2(t)]^T$ is the vector of probability amplitudes for finding the system in the basis state $|1\rangle$ and $|2\rangle$, respectively. Basis states are chosen as eigenstates of the field-free Hamiltonian of the system. Such basis is called \textit{diabatic} or \textit{bare state} basis. In the presence of time-varying electric fields the total Hamiltonian of the system is given in the rotating-wave approximation (RWA) \cite{Shirley1965, Shore2009} as

\begin{equation}
\mathbf{H}(t) = 
  \left( {\begin{array}{cc}
   0 & \frac{1}{2}\Omega(t) \\
   \frac{1}{2}\Omega(t) & \Delta(t) \\
  \end{array} } \right)
\label{eq:RWA}
\end{equation}
where $\Omega(t)=-\mathbf{d}_{12}\cdot\mathbf{\varepsilon}(t)$ is the \textit{Rabi} frequency \cite{gerry_knight_2004} associated with the electric field $\mathbf{\varepsilon}(t)$, which couples states $|1\rangle$ and $|2\rangle$ through the transition dipole moment vector $\mathbf{d}_{12}$. Atomic units are used throughout this paper. The dynamic detuning $\Delta(t)=S(t)-S_0$ is defined as the difference between the static detuning $S_0=\omega-\omega_0$ and the time-dependent Stark shift $S(t)$. The static detuning $S_0$ is the difference between the frequency $\omega$ of the electric field $\mathbf{\varepsilon}(t)$ and the \textit{Bohr} frequency $\omega_0$ (energetic separation) of states. With this definition, the resonance occurs at $\Delta(t)=0$. 

Time evolution of the probability amplitudes defined in eq. \ref{eq:wf} is described by the Schr{\"o}dinger equation

\begin{equation}
i\frac{d\mathbf{C}(t)}{dt}=\mathbf{H}(t)\mathbf{C}(t)
\label{eq:SE2}
\end{equation}
which, in this representation, is a system of first-order ordinary differential equations for complex amplitudes $c_1(t),c_2(t)$. No interaction with the environment and no other decoherence sources are assumed, as well as spontaneous emission is neglected. Under these conditions the total probability is conserved in the system, meaning that $|c_1(t)|^2+|c_2(t)|^2=1$ at all times $t$. 

\subsection*{The idea behind SCRAP}

Assume that the system is initially ($t=t_i$) in state $|1\rangle$, so that $P_1(t_i)=|c_1(t_i)|^2=1$ and $P_2(t_i)=|c_2(t_i)|^2=0$. The objective is to transfer the population completely from state $|1\rangle$ to state $|2\rangle$ at time $t_f > t_i$, i.e. $P_1(t_f)=|c_1(t_f)|^2=0$ and $P_2(t_f)=|c_2(t_f)|^2=1$. For realization of such complete population transfer between quantum states a robust protocol, called Stark-chirped rapid-adiabatic passage (SCRAP) \cite{Rickes2000,Rangelov2005,Shore2009,Vitanov2017}, can be used. This scheme relies on adiabatic time evolution of the state vector given in eq. \ref{eq:wf}. For this reason let us introduce the adiabatic approximation into the present problem. The Hamiltonian matrix in eq. \ref{eq:RWA} in the \textit{diabatic} representation can be diagonalized by means of a unitary transformation:

\begin{equation}
\mathbf{U}(t) = 
  \left( {\begin{array}{cc}
   \cos\Theta(t) & -\sin\Theta(t) \\
   \sin\Theta(t) & \cos\Theta(t) \\
  \end{array} } \right)
\label{eq:AD-DB}
\end{equation}
where $\tan 2\Theta(t)=\frac{\Omega(t)}{\Delta(t)}$.
This transformation results in a change of basis into the adiabatic basis: $\left(|\phi_-(t)\rangle,|\phi_+(t)\rangle\right)^T=\mathbf{U}(t)\left(|1\rangle,|2\rangle\right)^T$. The transformed Hamiltonian is given by

\begin{equation}
\mathbf{\tilde{H}}(t) =\mathbf{U}^T\mathbf{H}\mathbf{U}= 
  \left( {\begin{array}{cc}
   \epsilon_-(t) & -\dot{\Theta}(t) \\
   \dot{\Theta}(t)  &  \epsilon_+(t) \\
  \end{array} } \right)
\label{eq:HAD1}
\end{equation}
When the mixing angle $\Theta(t)$ varies sufficiently slowly with time and the separation between the \textit{adiabatic energies} $\epsilon_{\pm}(t)=\frac{1}{2}\Delta(t)\pm \frac{1}{2}\sqrt{\Delta(t)^2+\Omega(t)^2}$ is large enough, i.e. ($| \dot{\Theta}(t)| \ll | \epsilon_+(t)- \epsilon_-(t)|$), then the off-diagonal terms in eq. \ref{eq:HAD1} can be neglected and time evolution of the system is considered adiabatic \cite{Rangelov2005,Shore2009,Vitanov2017}. Under such conditions, the initial state of the system, which was assumed to be one of the adiabatic states $\phi_{+}(t_i)$ or $\phi_-(t_i)$, will remain unchanged (up to a phase factor) during the evolution, even though the individual \textit{diabatic} components of the adiabatic state evolve in time.   

If external electric fields are tailored so that at $t=t_i$ the mixing angle is $\Theta(t)=\pi/2$, then the adiabatic state reads $\phi_+(t_i)=|1\rangle$. This is achieved when the initial Rabi frequency $\Omega(t_i)$ is small relative to the magnitude of the detuning $\Delta(t_i)$, which is assumed negative. As time proceeds the detuning is allowed to pass through resonance ($\Delta(t_0)=0$) whilst keeping $\Omega(t_0)$ non-zero. At this point, the adiabatic state becomes an equally weighted mixture of diabatic states  $\phi_+(t_0)=\frac{1}{\sqrt{2}}|1\rangle+\frac{1}{\sqrt{2}}|2\rangle$. Finally, at time $t_f\gg t_0$ the Rabi frequency $\Omega(t_f)$ is expected to be small relative to the detuning $\Delta(t_f)$, which is expected to be large and positive. Such situation corresponds to mixing angle $\Theta(t_f)=0$ and adiabatic state $\phi_+(t_f)=|2\rangle$. If all changes to $\Omega(t)$ and $\Delta(t)$ were continuous and slow enough to satisfy the adiabatic condition, then the complete population transfer (CPT) is achieved between the diabatic states $|1\rangle \rightarrow |2\rangle$. The general sequence outlined above carries a name of rapid-adiabatic passage (RAP), and when the changes to the dynamic detuning $\Delta(t)$ are caused by chirping the separation of energy levels with an external electric field, then it is called Stark-chirped rapid-adiabatic passage (SCRAP) \cite{Rickes2000,Yatsenko1999,Rangelov2005,Shore2009,Vitanov2017}. 
Populations of diabatic states during SCRAP can be calculated from analytic expressions $P_2(t)=|c_2(t)|^2=\cos^2\Theta(t)$, $P_1(t)=|c_1(t)|^2=\sin^2\Theta(t)$.

In a typical SCRAP scenario two electric fields are considered: Stark pulse $\epsilon_s(t)$ and pump pulse $\epsilon_p(t)$. The role of the Stark pulse is to chirp the detuning $S(t)$ smoothly through resonance $\Delta(t_0)=0$ ($S(t_0)=S_0$) with the frequency of the pump pulse. The pump pulse raises the value of the Rabi frequency $\Omega(t)$ for the $|1\rangle \leftrightarrow |2\rangle$ transition and its frequency $\omega$ is slightly blue-detuned (greater) from the zero-field energy separation $\omega_0$ of the energy levels of the system. At this point all electric fields are assumed $\hat{z}$-linearly polarized.

In this work we are going to focus on finding time profiles for controllable pulses $\Omega(t)$ and $\Delta(t)$ such that the population transfer in the SCRAP sequence described above is optimal under given conditions and perturbations.

\subsection*{Bloch equations}
In a closed quantum system the time evolution of the state vector can be described equivalently by the Schr{\"o}dinger equation or by the Liouville equation \cite{gerry_knight_2004} $i\dot{\mathbf{\rho}}(t)=\left[\mathbf{H}(t),  \mathbf{\rho}\right]$, which operates with the density matrix 

\begin{equation}
\mathbf{\rho}(t) = 
  \left( {\begin{array}{cc}
   \rho_{11}(t) & \rho_{12}(t) \\
   \rho_{21}(t) & \rho_{22}(t) \\
  \end{array} } \right)
\label{eq:rho}
\end{equation}
where $\rho_{11}(t)=|c_1(t)|^2, \rho_{22}(t)=|c_2(t)|^2, \rho_{12}(t)=c^*_1(t)c_2(t), \rho_{21}(t)=c^*_2(t)c_1(t)$. A convenient way of representing the dynamics of a two-level system is given by transformation of the two complex components of the probability amplitude vector $\mathbf{C}(t)$  into the space of three real components. The normalisation of the total wavefunction makes one of the components of this probability vector dependent on the other three. Similarly, the density matrix $\mathbf{\rho}$ can be written in a general form in the following way: $\mathbf{\rho}(t)=\frac{1}{2}\left(\mathbf{1}_2+\mathbf{R}(t)\cdot\mathbf{\sigma}\right)$, where $\mathbf{\sigma}=(\sigma_1,\sigma_2,\sigma_3)$ is the vector of \textit{Pauli matrices} and $\mathbf{R}(t)=(R_1(t),R_2(t),R_3(t))$ is the \textit{Bloch vector} \cite{gerry_knight_2004}, which fully describes the state of the quantum system at time $t$. 

The time evolution of the Bloch vector in the \textit{Bloch representation} with the  Hamiltonian from eq. \ref{eq:RWA}  is given by the equation 

\begin{equation}
  \left( {\begin{array}{c}
   \dot{R}_1(t)  \\
   \dot{R}_2(t) \\
   \dot{R}_3(t) \\
  \end{array} } \right)=   \left( {\begin{array}{ccc}
   0 & -\Delta(t) & 0 \\
   -\Delta(t) &  0 & -\Omega(t) \\
   0 & \Omega(t) & 0 \\
  \end{array} } \right)  \left( {\begin{array}{c}
   R_1(t)  \\
   R_2(t) \\
   R_3(t) \\
  \end{array} } \right)
\label{eq:bloch}
\end{equation}
subject to state constraints $||\vec{R}||=1$, which is equivalent to requiring the total wavefunction to be normalized at all times: $|c_1(t)|^2+|c_2(t)|^2=1, \; \forall t$. We assumed, without loss of generality that the Rabi frequency is real ($\Omega^*=\Omega$) \cite{Shore2009}. In a shorthand notation the above equation can be written as $\dot{\mathbf{R}}(t)=\mathbf{M}(t) \mathbf{R}(t)$. The conservation of the total probability means that the quantum dynamics is realized on the three-dimensional sphere spanned by the Bloch vector $\vec{R}$. The state vector of the system can be then parametrized by two angles $(\theta,\phi)$ as follows
\begin{equation}
|\Psi(t)\rangle = \cos\theta(t)e^{i\phi}|1\rangle+\sin\theta(t)e^{-i\phi}|2\rangle
\label{eq:blochstate}
\end{equation}
and the complete population transfer corresponds to any path starting from the south pole of the Bloch sphere ($|c_1(t_i)|^2=1, \mathbf{R}(t_i)=(0,0,-1)^T$) and finishing at the north pole ($|c_2(t_f)|^2=1, \mathbf{R}(t_f)=(0,0,1)^T$), which corresponds to change in the azimuthal angle $\theta$ from $\pi$ to $0$.

\subsection*{Optimal control theory for SCRAP}

For the dynamics given by the Bloch equations:
\begin{equation}
\dot{\mathbf{R}}(t)=\mathbf{M}(t) \mathbf{R}(t)
\label{eq:Bloch}
\end{equation}
with the initial $\mathbf{R}(t_i)=(0,0,-1)^T$ and the target condition $\mathbf{R}(t_f)=(0,0,1)^T$, the goal is to minimize a cost functional $P[\vec{\alpha}(t)]$. This cost functional depends on a vector of control functions $\vec{\alpha}$. With this minimization problem and with the dynamics given in eq. \ref{eq:Bloch} a pseudo-Hamiltonian is associated \cite{jurdjevic,Bonnard2012}:

\begin{equation}
H(\mathbf{R},\mathbf{p},t;\vec{\alpha}(t))=\mathbf{f}(t)\cdot \mathbf{p}(t)+p_0r(\vec{\alpha}(t),t)
\label{eq:pseudoHam}
\end{equation}
where $\mathbf{f}(t)=\mathbf{M}(t) \mathbf{R}(t)$ is the right hand side of the Bloch equations, $\mathbf{p}(t)$ is the vector of \textit{costate} functions, which correspond to the generalized momenta in the classical Hamilton dynamics and $r(t)$ is related to the cost function by the formula $P[\vec{\alpha}]=\int_{t_i}^{t_f}r(\vec{\alpha}(t),t)dt$. $p_0$ is a real normalisation constant for $r(\vec{\alpha}(t),t)$. The Pontryagin Maximum Principle (PMP) states that maximization of the Hamiltonian in eq. \ref{eq:pseudoHam} with respect to the control functions $\vec{\alpha}(t)$ provides the optimal set of parameters $\vec{\alpha}^*(t)$ for the dynamics given in eq. \ref{eq:Bloch} and minimize the cost functional $P[\vec{\alpha}]$ \cite{jurdjevic,Bonnard2012,Boscain2002,Assmat2012,VanDamme2017}. The conditions for the maximization of the pseudo-Hamiltonian are written as
\begin{small}
\begin{equation}
\begin{split}
H(\mathbf{R},\mathbf{p},t;\vec{\alpha}^*(t))= \max_{\vec{\alpha}\in \mathcal{CC}} \lbrace H(\mathbf{R},\mathbf{p},t;\vec{\alpha}(t))| C(\mathbf{R}(t),\vec{\alpha}(t))=0\rbrace\\
\vec{\nabla}_{\vec{\alpha}} H(\mathbf{R},\mathbf{p},t)=\lambda(t)\nabla_{\alpha}C(\mathbf{R}(t),\vec{\alpha}(t))+\sum_{j=1}^{N_c}\mu_j(t)\nabla_{\alpha}h_j(\vec{\alpha})
\end{split}
\label{eq:optcont}
\end{equation}
\end{small}
where $C(\mathbf{R}(t),\vec{\alpha}(t))=\mathbf{f}(t)\cdot\nabla g(\mathbf{R})$, and $\lambda(t), \lbrace\mu_j\rbrace_{j=1,2,...,N_c}$ are Lagrange multipliers associated with constraints on the state vector and the control fields. The state constraints $SC$ set is defined as $SC=\lbrace \mathcal{R}\in\mathcal{R}^3 | g\left(\mathbf{R}(t)\right)=R_1^2(t)+R_2^2(t)+R_3^2(t)-1=0\rbrace$ and the control constraints $CC$ set is defined as $CC=\lbrace \vec{\alpha}\in\mathcal{R} | \mathbf{h}(\vec{\alpha})\leq 0, \dim(\mathbf{h})=N_c \rbrace$. 
The optimal trajectories $(\mathbf{R}(t),\mathbf{p}(t))$ can be calculated by solving  the equations of motion with the optimized control parameters $\vec{\alpha}^*(t)$:

\begin{equation}
\begin{split}
\dot{\mathbf{R}}(t)=\vec{\nabla}_p H(\mathbf{R},\mathbf{p},t;\vec{\alpha}^*(t)) \\
\dot{\mathbf{p}}(t)=-\vec{\nabla}_R H(\mathbf{R},\mathbf{p},t;\vec{\alpha}^*(t))+\lambda(t)\nabla_R  C(\mathbf{R}(t),\vec{\alpha}(t))
\end{split}
\label{eq:optdyn}
\end{equation}

For the dynamics of a quantum closed system, where the Bloch vector is normalized to unity at all times, the $C(\mathbf{R}(t),\vec{\alpha}(t))$ function vanishes by identity ($C\equiv 0$). 

Given the Bloch equations \ref{eq:bloch}, the PMP pseudo-Hamiltonian can be written explicitly as:
\begin{equation}
H(\mathbf{R},\mathbf{p},t;\vec{\alpha}(t))=l_1(t)\Omega(t)+l_3(t)\Delta(t)+p_0r(\vec{\alpha}(t),t)
\label{eq:Ham1}
\end{equation}
where $l_1(t)=R_2(t)p_3(t)-R_3(t)p_2(t)$ and $l_3(t)=R_1(t)p_2(t)-R_2(t)p_1(t)$ can be regarded as the components of the angular momentum vector $\mathbf{l}=(l_1,l_2,l_3)=\mathbf{R}\times \mathbf{p}$, which is associated with motion on the Bloch sphere. Here we intend to minimize the total energy of the Stark pulse ($\Delta(t)$) and the pump pulse ($\Omega(t)$) with simultaneous condition for the complete population transfer between states $|1\rangle$ and $|2\rangle$. Thus, the cost functional is given by $P[\vec{\alpha}(t)]=\int_{t_i}^{t_f}\left(\Delta(t)^2+\Omega(t)^2\right) dt$, so that $r(\vec{\alpha}(t),t)=\Delta(t)^2+\Omega(t)^2$. Constant $p_0$ can be normalized to the commonly used value $p_0=-1/2$ \cite{VanDamme2017}. The control fields vector is two-dimensional $\vec{\alpha}(t)=(\Delta(t),\Omega(t))$.

The equations presented above concern the case of control pulses, which are independent of position coordinates. Let us now assume that the control parameters, as well as the state vectors, depend on position $z$, meaning that we have an ensemble of systems (molecules) placed in spatially inhomogeneous electric fields in 1-dimension $z$. Each system experiences different electric field strengths and chirp rates. 
Accordingly, the Rabi frequency $\Omega(t)$ and the dynamic detuning $\Delta(t)$ depend on position $z$ and are subject to a space- and time- varying perturbation defined by functions $K_1(z),K_2(z),f_1(z),f_2(z)$. Under such circumstances the general control parameters can be rewritten as

\begin{equation}
\begin{split}
\Omega(z,t)=(1+f_1(t)K_1(z))\Omega(t)\\
\Delta(z,t)=(1+f_2(t)K_2(z))\Delta(t)
\end{split}
\label{eq:controlparams}
\end{equation} 
Then the cost functional becomes a sum over all positions $z$ in a given range $[z_{min},z_{max}]$: $P[\vec{\alpha}(t)]=\int_{z_{min}}^{z_{max}}dz\int_{t_i}^{t_f}r(\vec{\alpha}(z,t),t)dt$. The situation of such inhomogeneous perturbing electric field is discussed in section \ref{sec:general}.

\section{Adiabatic dynamics}
\label{sec:gauss} 
This section focuses on the adiabatic quantum dynamics of a two-level system. In such case, an analytic expression for the time profile of the population of diabatic states is given by equation \ref{eq:AD-DB}. Fixed functional forms for the Stark and the pump pulse are used, and their parameters are optimized. This approach corresponds to the problem of finding maxima of a multivariate function. More general cases, which require involving the PMP, are discussed in section \ref{sec:general}.

\subsection*{Gaussian Stark and pump pulses}

Gaussian profiles for the Stark pulse the and pump pulse were chosen: 

\begin{equation}
\begin{split}
\Delta(t) = -S_0 + 
   \frac{\Delta_0}{\sqrt{2 \pi}\sigma_s}\exp\left(-\frac{(t - t_s)^2}{2\sigma_s^2}\right) \\
   \Omega(t) = \frac{\Omega_0}{\sqrt{2 \pi}\sigma_p}\exp\left(-\frac{(t - t_p)^2}{2 \sigma_p^2}\right)
\end{split}
\label{eq:pulses}
\end{equation}
where $S_0=\omega-\omega_0$ is positive static detuning. Normalized Gaussian pulses were scaled by amplitude factors $\Delta_0$ and $\Omega_0$, respectively. Here $t_s,t_p$ are the centres of pulses in time domain and $\sigma_s,\sigma_p$ are respective widths of the Stark and the pump pulse. Such arrangements of electric fields corresponds to a typical SCRAP situation. Without significant loss of generality we can set the parameters of the pump pulse constant and adjust only the width $\sigma_s$ and the centre $t_s$ of the Stark pulse.
 
We aim at finding a set of parameters $(t_s,\sigma_s)$, which ensure complete population transfer between the initial state and the target state, by means of the rapid-adiabatic passage. First, it is necessary to chose an appropriate measure of the efficiency of the population transfer process. The population of the target state $P_{2}(T;t_s,\sigma_s)$ evaluated at time $T$ can serve as such a measure.
The choice of the 'probe time' $T$ is somewhat arbitrary. Here the probe time was chosen based on the parameters of the pump pulse: $T=t_p+3\sigma_p$. The value of the probe time depends on the range of the pump pulse, which is necessary for adiabatic evolution of the state vector of the system. Thus, $T$ is located in just prior to the time region, where the adiabatic evolution is expected to break down, i.e. where the amplitude of the pump pulse becomes small.

$P_{2}(T;t_s,\sigma_s)=\cos^2\left(\Theta\left(T;t_s,\sigma_s\right)\right)$ is a parametric function of the width and centre of the Stark pulse. From now on, we use reduced units defined with respect to the pump pulse parameters: $\tau = (t_s-t_p)/t_p$ and $\sigma=\sigma_s/\sigma_p$.

\begin{figure}
\includegraphics[width=8.6cm]{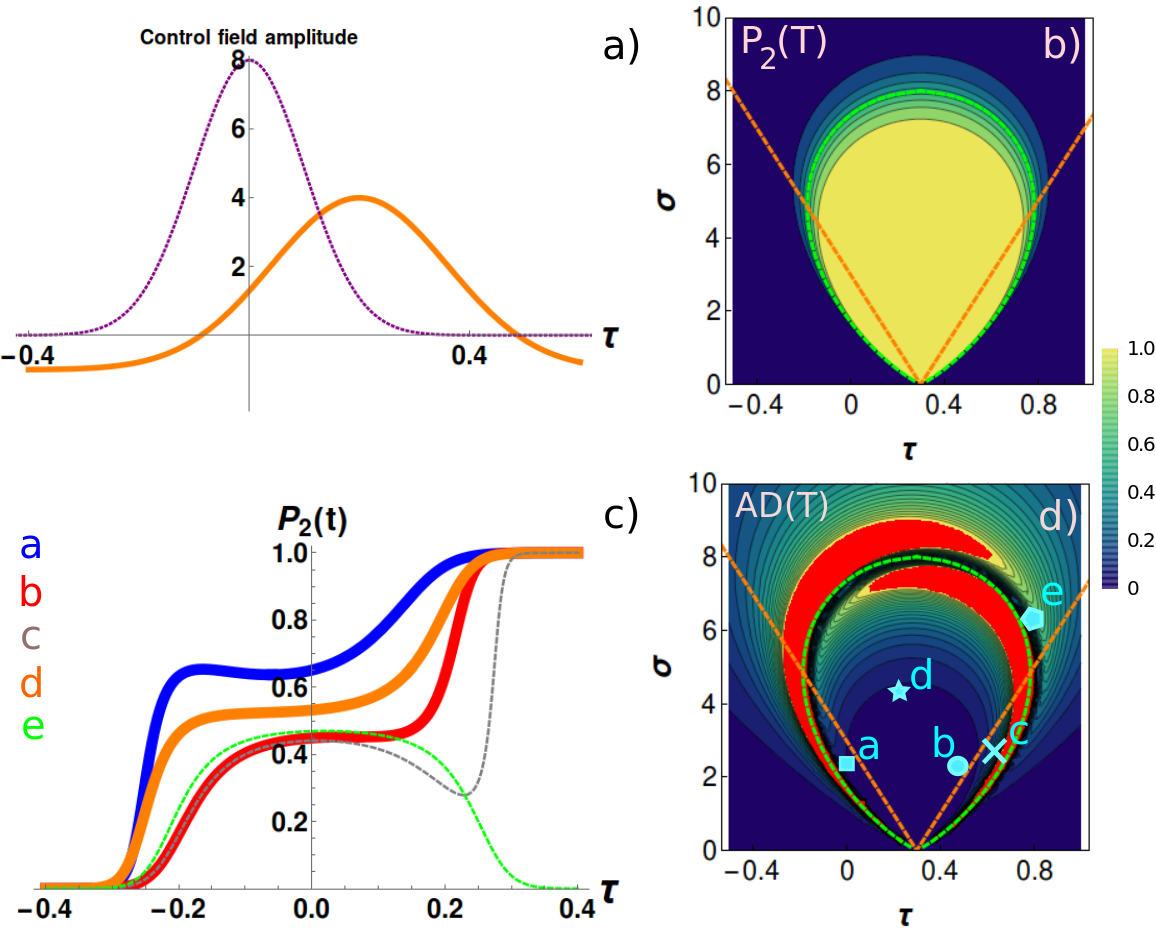}
\caption{Stark-chirp rapid-adiabatic passage for Gaussian pulses in the adiabatic approximation. The following parameters for pulses were used: $S_0=1.0, \Omega_0=100.0, \Delta_0=100.0, \sigma_p=5.0, t_p=50.0$. The controllable parameters are the reduced width $\sigma$ and the reduced centre $\tau$ of the Stark pulse. a) Time profiles for the pump pulse (purple dashed line) and the Stark pulse;  b) The efficiency map for the population transfer with the Gaussian pulses used. See text for further explanation; c) Time profiles for the population of the target state plotted for several choices of the reduced parameters (indicated with letters a-e); d) Adiabaticity function $AD(T)$ plotted for the reduced parameters at probe time $T=t_p+3\sigma_p$.  }
\label{fig:GGpulses}
\end{figure}

A contour plot of the efficiency measure $P_{2}$ as a function of the reduced Stark pulse parameters is displayed in Figure \ref{fig:GGpulses}b. This graph shows that there exists a specific region in the Stark pulse parameters' space for which the population transfer should be nearly complete (marked with the yellow colour in Figure \ref{fig:GGpulses}c).

Local maxima of the $P_{2}$ function can be calculated by solving the system of two equations: 
\begin{equation}
\frac{\partial P_{2}(T;\tau,\sigma)}{\partial \sigma}=0,  \; \frac{\partial P_{2}(T;\tau,\sigma)}{\partial \tau}=0 
\label{eq:GGcond}
\end{equation}
Derivation of analytic solutions to this system of equations is given in appendix A. Each equation yields a pair of curves $\tau^{\pm}(\sigma)$, which are depicted in Figure \ref{fig:GGpulses}b and \ref{fig:GGpulses}d with green and orange dashed lines. Each single curve leads over saddle points in the parameters' landscape.  Intersections of these saddle curves indicate candidate points for local maxima of the efficiency function $P_2$. Three intersection points can be found: $(\tau,\sigma)=(0.3,0),(-0.18,4.8),(0.78,4.8)$.

Analytic functions $\tau^{\pm}(\sigma)$, which satisfy conditions from eq. \ref{eq:GGcond} define a candidate region in the parameters' space for the optimal population transfer. Equations \ref{eq:GGcond} provide only the necessary condition for the existence of maximum. The critical point $(\tau,\sigma)=(0.3,0)$ with zero Stark pulse width is rejected as non-physical. The other two points are are symmetrical with respect to the chosen probe time $T=0.3$.

Because we are operating within the model of the adiabatic time evolution of the state vector of the system, it is necessary to find a region in the space of the Stark pulse parameters, which satisfy the adiabatic condition: $|\Omega(t)\dot{\Delta}(t)-\dot{\Omega(t)}\Delta(t)|\ll 2\left(\Omega(t)^2+\Delta(t)^2\right)^{\frac{3}{2}}$ \cite{Rangelov2005,Shore2009}. For this purpose let us define the \textit{adiabaticity function} $AD(t;\tau,\sigma)=\frac{|\Omega(t)\dot{\Delta}(t)-\dot{\Omega(t)}\Delta(t)|}{2\left(\Omega(t)^2+\Delta(t)^2\right)^{\frac{3}{2}}}$, which is plotted for $t=T$ in Figure \ref{fig:GGpulses}d. 

Values of $AD(T;t_s,\sigma_s)$ significantly smaller than 1 suggest adiabatic evolution of the state vector during SCRAP. Red regions in the contour plot \ref{fig:GGpulses}d denote highly non-adiabatic time evolution.

By overlapping the adiabaticity function with the efficiency function from Figure \ref{fig:GGpulses}b, a region in the $(\tau,\sigma)$ space can be found, for which the time evolution is expected to be optimal and adiabatic. To illustrate this, in Figure \ref{fig:GGpulses}c time profiles for the population of the target state $P_2(t)$ are depicted for several choices of the Stark pulse parameters, which are indicated with points and letters (a-e) in Figure \ref{fig:GGpulses}d. For points located within the adiabatic and the optimal region, the population transfer is complete (profiles a,b and d). For a point from outside the optimal region, marked with "e" and cyan pentagon in Figure \ref{fig:GGpulses}d, the time evolution is no longer complete. Thus even though the analytic optimal curves $\tau^{\pm}(\sigma)$ suggest the near optimal population transfer, only the region with a relatively small width of the stark Pulse can be considered feasible in the present model. 
This is due to limited width of the pump pulse, which ensures avoided crossing between adiabatic surfaces of energies of states and guarantees adiabatic evolution. For the chosen point "e" the Stark pulse chirps through resonance at times when the pump pulse has already nearly vanished (the far right tail of the pump pulse). For this reason the population cannot be fully transferred between the initial and the target state. To conclude, even if the necessary condition, suggested in Figure \ref{fig:GGpulses}b, for complete population transfer is satisfied, the adiabatic condition depicted in Figure \ref{fig:GGpulses}d can eliminate certain pulse shapes. But because the evolution is diabatic near the second \textit{mute} resonance, the full dynamical model, which accounts for adiabatic and diabatic evolution, is more adequate.    

\subsection{Optimization of the full Bloch dynamics for Gaussian Stark and Gaussian pump pulses in SCRAP}
 
\begin{figure}
\includegraphics[width=8.6cm]{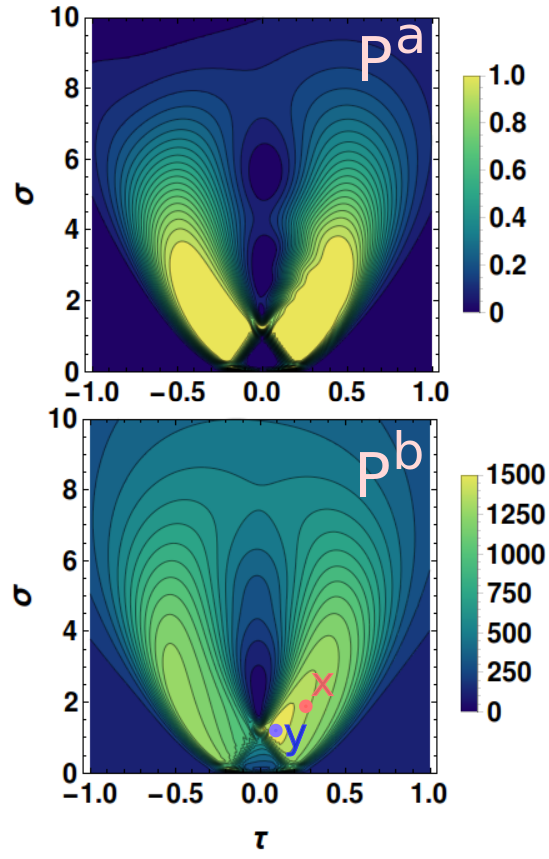}
\caption{Efficiency maps ($P^{a}_2(T)$-top graph, $P^{b}_2(T)$-bottom graph) for SCRAP calculated by solving full Bloch equations given in \ref{eq:bloch}. Parameters used were identical as for generation of Figure \ref{fig:GGpulses}: $S_0=1.0, \Omega_0=100.0, \Delta_0=100.0, \sigma_p=5.0, t_p=50.0$. The controllable parameters are the reduced width $\sigma=\sigma_s/\sigma_p$ and reduced time $\tau=(t_s-t_p)/t_p$ of the Stark pulse. }
\label{fig:GGmaps}
\end{figure}

The efficiency measure $P_2$ used in the adiabatic approximation can be generalized to the case of the full quantum dynamics on the Bloch sphere: $P^{a}_2(T)=|c_2(T)|^2=\frac{1}{2}+\frac{1}{2}R_3(T)$. Alternatively one can choose an integral measure  $P^{b}_2(T)= \int_{t_i}^T |c_2(T)|^2dt= \int_{t_i}^T \left(\frac{1}{2}+\frac{1}{2}R_3(t)\right)dt$. The efficiency measures defined above are plotted in Figure \ref{fig:GGmaps} as a function of the reduced parameters $(\tau,\sigma)$. The maps were calculated by solving the Bloch equations given in eq. \ref{eq:bloch} with the Stark and the pump pulses defined in eq. \ref{eq:pulses}. All other parameters were identical as in the adiabatic case.

%Finding optimal solutions
By requiring the gradient of the efficiency measures to vanish $\nabla_{\tau,\sigma}P^{a/b}_2(T)=0$ we find maxima of these efficiency maps in the Stark pulse parameters' space. For $P^{a}$ we have $(\tau^*,\sigma^*)=(0.3,2.0)$ and for $P^{b}$ we have $(\tau^*,\sigma^*)=(0.07,1.2)$. Values of the optimal parameters for both measures are similar, as expected from similar efficiency measures. 

\begin{figure*}
\includegraphics[width=\textwidth]{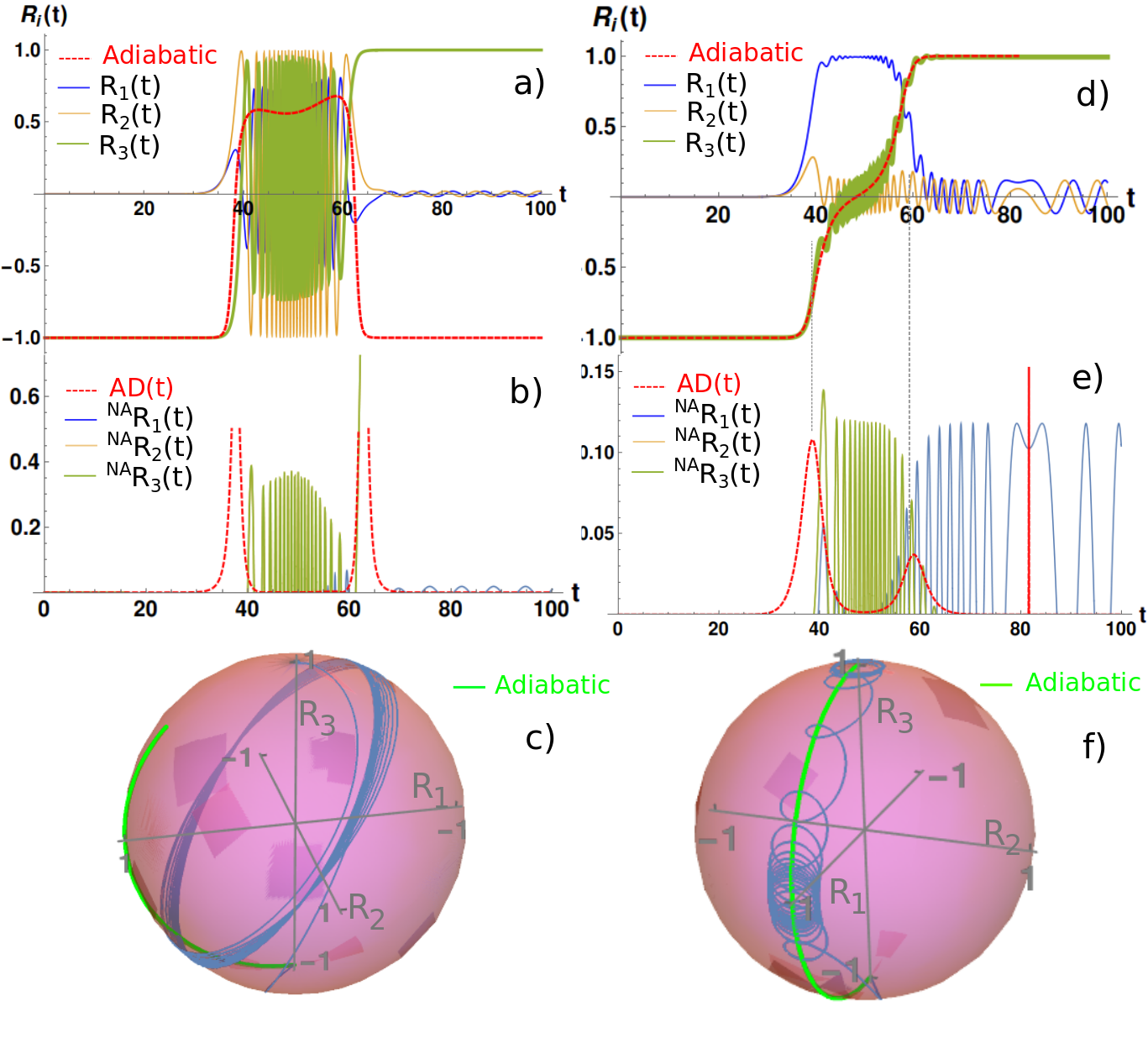}
\caption{The Bloch vector dynamics during SCRAP for two different Stark pulse shapes. The graphs on the left hand side correspond to the global maximum of the efficiency map $P^a$ marked with "y" in Figure \ref{fig:GGmaps}, whereas the graphs on the right hand side correspond to the point marked with "x" in Figure \ref{fig:GGmaps}. Sub-figures a) and d) display time profiles for components of the Bloch vector. The dashed red line stands for the adiabatic evolution in the Bloch picture. Sub-figures b) and e) depict non-adiabatic contribution to the time evolution of the Bloch vector components during SCRAP. Red dashed line represents the adiabacity function $AD(t)$ defined as $AD(t)=\frac{|\Omega(t)\dot{\Delta}(t)-\dot{\Omega(t)}\Delta(t)|}{2\left(\Omega(t)^2+\Delta(t)^2\right)^{\frac{3}{2}}}$. Sub-figures c) and f) show trajectories followed by the Bloch vector on the Bloch sphere. In dashed green the trajectories resulting from the adiabatic approximation are also displayed. In the calculations $t_i=0$, $t_f=100$. }
\label{fig:NABC}
\end{figure*}

Figures \ref{fig:NABC}(d-f) refer to the set of parameters marked in Figure \ref{fig:GGmaps} with the red "\textit{x}" $(\tau,\sigma)=(0.3,2)$, a point which belongs to the optimal region for the measure $P^a$ and a nearly optimal region for the measure $P^b$. Figures \ref{fig:NABC}(a-c) refer to the global maximum of the efficiency map $P^b$ (marked in Figure \ref{fig:GGmaps} with the blue "\textit{y}" $(\tau,\sigma)=(0.07,1.2)$).

Figures \ref{fig:NABC}a and \ref{fig:NABC}d show time profiles for the components of the Bloch vector during SCRAP. The $R_3(t)$ component of the Bloch vector, which tracks the population flow between the initial and the target state follows a complete path from the initial value -1 at $t=t_i$ to the final value +1 at $t=t_f$. 
Only the $R_3(t)$ profile in Figure \ref{fig:NABC}d resembles the time profile calculated in the adiabatic approximation. For both used sets of the Stark pulse parameters the full dynamics calculations show some oscillations in $R_3(t)$, which arise due to non-adiabaticity of the state vector evolution. This non-adiabatic behaviour can be  quantified with the aid of non-adiabatic Bloch correction (NABC) vector, defined as $\vec{R}^{NA}(t)=\vec{R}(t)-\vec{R}^{AD}(t)$. This vector informs about deviations from the adiabatic dynamics. Here $\vec{R}^{AD}=(\sin2\Theta(t),0,-\cos2\Theta(t))^T$. Figures \ref{fig:NABC}b,e display the components of the NABC as a function of time, together with the adiabaticity function plotted with dashed red lines. 

Rises in the value of the adiabaticity function correlate with onsets of non-adiabatic behaviour of the Bloch vector. When the non-adiabatic contribution to the dynamics is small, as seen in Figure \ref{fig:NABC}e, the adiabatic analytic formula for the population transfer agree well with the full numerical solutions to the Bloch equations. The non-adiabatic contribution to the $R_3$ and $R_1$ components of the Bloch vector are no greater than 20\%, suggesting that the dynamics is to a good extend adiabatic.
On the other hand, rapid oscillations of the Bloch vector components visible in Figure \ref{fig:NABC}a lead to high values of the adiabaticity function, which results in highly non-adiabatic time evolution. In such case the adiabatic formula for the population transfer show the complete population return, whereas the full numerical calculations suggest the complete population transfer. The discrepancy between the adiabatic and numerical profiles begins when the components of the Bloch vector start oscillate rapidly. This can be seen in Figure \ref{fig:NABC}c as multiple revolutions around the Bloch sphere  before reaching the north pole. In contrast, in Figure \ref{fig:NABC}f, the trajectory followed by the Bloch vector only weakly oscillate around the adiabatic path, which means that the time evolution of the state vector is not 'kicked' away from the adiabatic regime. The magnitude of the extra precession around the $R_3$ and the $R_1$ axis informs about the non-adiabatic contribution to the dynamics.

Thus we may conclude that the "y" pair of the globally optimal parameters generate highly non-adiabatic dynamics, which may be difficult to control. On the other hand, the point "x" located in the proximity of the global maximum in Figure \ref{fig:GGmaps} generates a smooth and nearly complete passage from the initial to the target state.

%Critique
The approach in which one finds maxima of the efficiency map should be thus criticized. No particular constraints have been imposed on the total energy or duration of the pulses. 
The optimal control theory discussed in the next section provides a way for designing pulses, which ensure an adiabatic evolution during SCRAP.

\section{Optimal control theory for SCRAP}
\label{sec:general}

\subsection*{Fixed pump pulse, Stark pulse optimized}

In this section we consider a special case when the pump pulse has a fixed functional form, while the Stark pulse is optimized with the use of the Pontryagin maximum principle. This type of problem in the optimal control theory is called the \textit{fixed endpoints problem} \cite{jurdjevic,Bonnard2012}, where the endpoints are given by the initial and target conditions for the Bloch vector ($\mathbf{R}(t_i)=(0,0,-1)^T, \mathbf{R}(t_f)=(0,0,1)^T$). The necessary condition for the maximization of the Hamiltonian is the vanishing derivative with respect to the Stark control field:

\begin{equation}
\frac{\partial H}{\partial \Delta} =0 \Longleftrightarrow l_3(t)-\Delta(t)=0
\label{eq:dHddelta}
\end{equation}
so that the optimal control for Stark pulse is given by the third component of the angular momentum $\Delta(t)\equiv \Delta^*(t)=l_3(t)$. The fixed pump pulse is chosen Gaussian-shaped as given in eq. \ref{eq:pulses}, with the width $\sigma_p = 5.0$,  the centre $t_p=50.0$ and the area $\Omega_0=100.0$. 

Solutions to the Hamilton equations (\ref{eq:optdyn}) with optimal the control Stark field yield trajectories followed by the Bloch vector $\mathbf{R}$ and the costate functions $\mathbf{p}$. Figure \ref{fig:gpmp}a shows time profiles for components of the Bloch vector during SCRAP with the optimized Stark pulse only and the Gaussian-shaped pump pulse. 
Figure \ref{fig:gpmp}b displays the $l_3(t)$ component of the angular momentum, which at the same time is the optimized Stark field. It is seen that the optimal pulse rapidly oscillates, passing through multiple resonances, a characteristics that renders the present scenario as largely impractical.  The Bloch vector makes a number of spurious revolutions around the Bloch sphere before reaching the north Pole, as shown in Figure \ref{fig:gpmp}b.

For this reason the results obtained with the Pontryagin maximum principle should be considered with caution, especially when there is not enough constraints imposed on the shape of the control fields. An additional condition is needed for control fields to achieve a more stable dynamics.

\subsection{The Stark and the pump pulse optimized}
\begin{figure*}
\includegraphics[width=\textwidth]{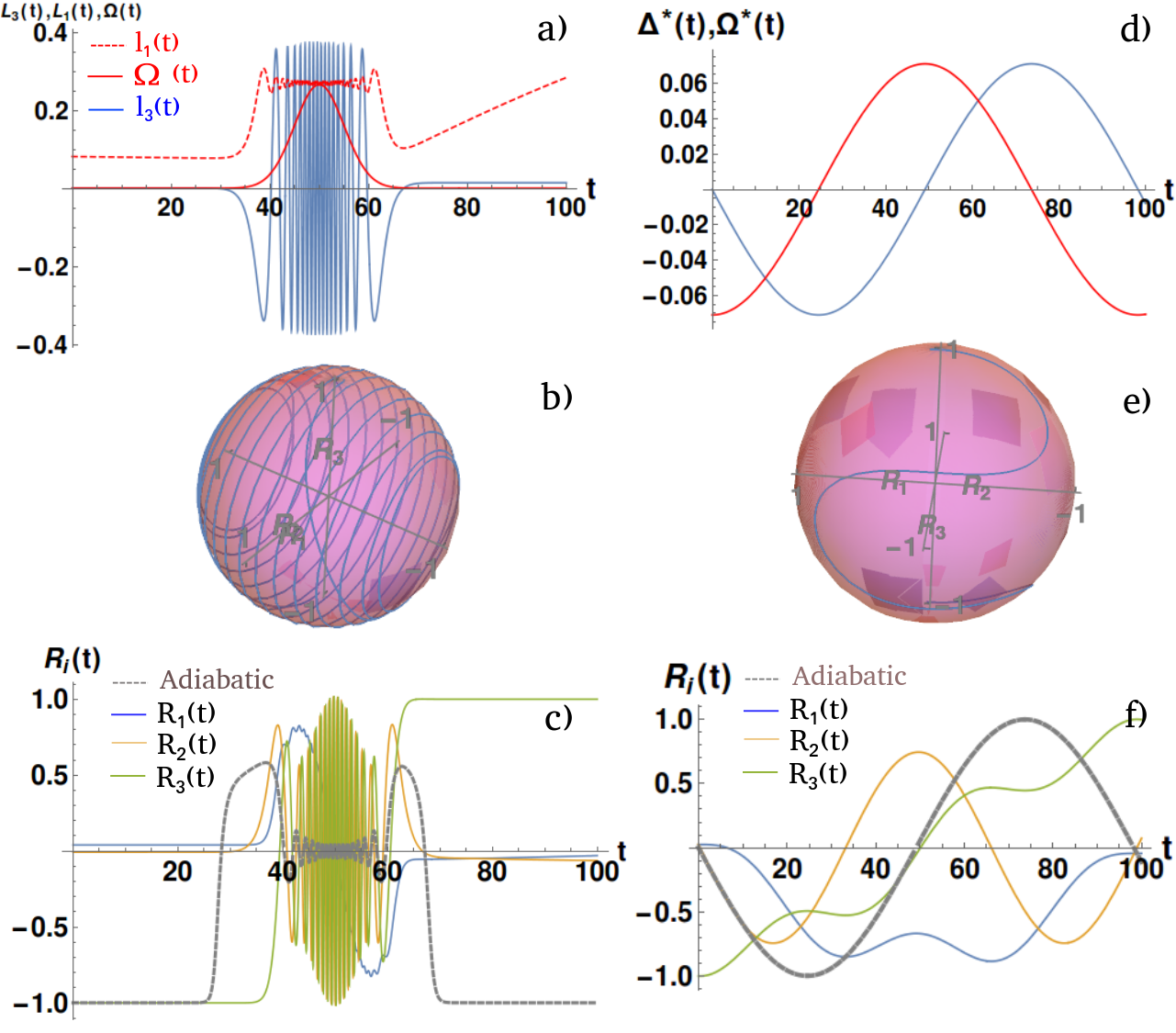}
\caption{Results of quantum dynamics calculations with optimal control fields for SCRAP. The left column corresponds to optimization of the Stark pulse only, with fixed Gaussian profile for the pump pulse. The right column shows results for optimization of both control fields. a),d) represent the optimized control fields: the Stark pulse and the pump pulse. a) also depicts with red dashed line the $l_1(t)$ component of the angular momentum. b),e) shows trajectories of the Bloch vector on the Bloch sphere. c),f) displays time profiles for components of the Bloch vector for optimal control fields. }
\label{fig:gpmp}
\end{figure*}

In this section we consider both the Stark pulse and the pump pulse as control fields subject to optimization. In such case there are two equations to solve for the optimal control fields:

\begin{equation}
\begin{split}
\frac{\partial H}{\partial \Delta} =0 \Longleftrightarrow l_3(t)-\Delta(t)=0\\
\frac{\partial H}{\partial \Omega} =0 \Longleftrightarrow l_1(t)-\Omega(t)=0
\end{split}
\label{eq:dHddo}
\end{equation}
The equations of motion for the Bloch and the costate vector are then written as:

\begin{equation}
\begin{split}
\dot{\mathbf{R}}(t)=\mathbf{l}(t)\times \mathbf{R}(t) \\
\dot{\mathbf{p}}(t)=\mathbf{l}(t)\times \mathbf{p}(t)
\end{split}
\label{eq:optdyn1}
\end{equation}
where $\mathbf{l}=(l_1(t),0,l_3(t))^T$ is the vector of the angular momentum defined in eq. \ref{eq:Ham1}. Mapping $t \mapsto H(\mathbf{R},\mathbf{p},t;\mathbf{\alpha}^*(t))$ is constant and equal to 0, as expected for optimal control fields. 
Another constant of motion is the square of the angular momentum $l^2=l^2_1(t)+l^2_2(t)+l^2_3(t)$. Figures \ref{fig:gpmp}(d-f) show results of quantum dynamics calculations with the optimal control fields $\Omega^*(t)=l_1(t),\; \Delta^*(t)=l_3(t)$.

Optimization of both the Stark and the pump pulse generates a rather smooth time profile for the population transfer curve ($R_3(t)$), as visible in Figure \ref{fig:gpmp}f, in contrast to the case when only the Stark pulse is optimized (Figure \ref{fig:gpmp}c). The optimal fields yield a 'tennis ball'-like trajectory on the Bloch sphere (shown in Figure \ref{fig:gpmp}e), which also satisfy the adiabatic condition, as expected from the rapid-adiabatic passage procedure. The bell-like shape of the optimized pump pulse and the smooth sine-like optimal Stark-chirp suggest that optimal control fields are in this case also rather intuitive.   

The remaining question about robustness of the optimal control fields to perturbations from the environment is discussed in the next section.

\subsection*{spatially inhomogeneous perturbation}
In this section we consider a 1-dimensional spatially inhomogeneous perturbation to the two-level quantum system, which affects the chirp rate in the Stark pulse. An example of such situation could be the space inside the Stark decelerator \cite{Bethlem2002PRL,Kupper2006} , where molecules experience strong inhomogeneities along the longitudinal direction from electric fields generated on metal rods. Each system (molecule) located at different $z$-position will experience slightly different chirp rate for a given Stark pulse. The aim is to find the shape of the Stark pulse, which guarantees an optimal population transfer in the SCRAP procedure for all systems (molecules) along a given $z$ region ($z\in [z_{min},z_{max}]$). In such case, the cost functional for the SCRAP process is the sum of contributions from all points along $z$:

\begin{equation}
P[\vec{\alpha}(t)]=\int_{t_i}^{t_f}dt\int_{z_{min}}^{z_{max}}r(\vec{\alpha}(z,t),t)dz
\label{eq:zdepmeasure}
\end{equation}

The amplitude of the Stark pulse experienced by the system is assumed to depend on position $z$ in the following way $
\Delta(z,t)=(1+K(z))\Delta(t)$ and the pump pulse is assumed to be unaffected by the spatial inhomogeneities. As a generic case let us assume that $K(z)=k(z-z_{min})$ is a linear function of $z$ in the region of interest ($z\in [z_{min},z_{max}]$). Analytic integration over $z$ of the expression for the total energy of the pulses $r(\mathbf{\alpha}(z,t),t)=\Delta^2(z,t)+\Omega^2(z,t)$ gives

\begin{equation}
\int_{z_{min}}^{z_{max}}r(\vec{\alpha}(z,t),t)dz=\frac{(1+kZ)^3-1}{3k}\Delta(t)^2-Z \Omega(t)^2
\label{eq:zdepmeasure1}
\end{equation}
yielding optimal control fields
\begin{equation}
\begin{split}
\Delta^*(t)=\beta l_3(t) \\
\Omega^*(t)=Z l_1(t)
\end{split}
\label{eq:zdepopt}
\end{equation}
where $Z = z_{max}-z_{min}$ and $\beta = \frac{3k}{1-(1+kZ)^3}$. In calculations $Z=1$ and $k=0.01$, which corresponds to a weak perturbation to the Stark chirp (less than 10\%).

\begin{figure*}
\includegraphics[width=\textwidth]{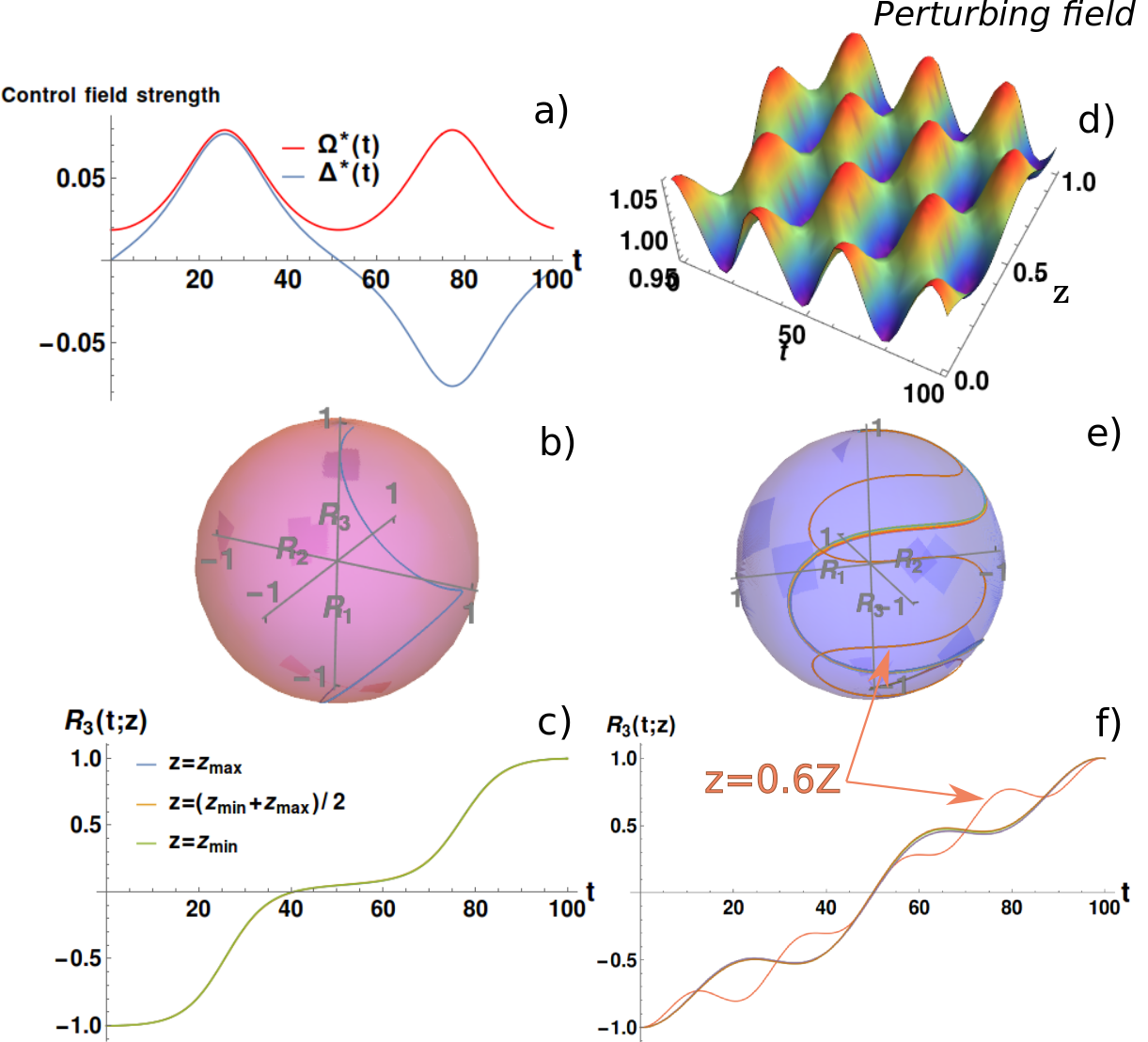}
\caption{Results of quantum dynamics calculations with optimal control fields for SCRAP with linear $z$-dependent perturbing electric field (a-c); $z$- and $t$-dependent perturbing electric field (d-f). a) Time profiles for the optimal Stark and pump pulses; b) Trajectory on the Bloch sphere for optimized SCRAP; c) Time profiles for the $R_3$ component of the Bloch vector for optimal control fields evaluated at $z$: $z_{min}, (z_{max}+z_{min})/2$ and $z_{max}$;  d) plot of the time and position dependent perturbation for $A=0.05, w=20, k=10$; e) several trajectories are plotted for different choices of the perturbing field parameters;  f) Time profiles for the $R_3$ component of the Bloch vector for optimal control fields evaluated at $z$: $z_{min}$ to $z_{max}$ with $0.1Z$ increment.   }
\label{fig:zdep}
\end{figure*}

All solutions $\mathbf{R}(t;z)$ depend parametrically on the position $z$. Figure \ref{fig:zdep} shows optimal control fields, the Bloch sphere trajectories and the time profile for the $R_3(t;z)$ component of the Bloch vector for three choices of $z$: $z_{min}, (z_{max}+z_{min})/2$ and $z_{max}$.

Solutions to Hamilton equations are sensitive to the value of the $\beta$ parameter.
Figure \ref{fig:zdep}a shows a double maximum shaped pump pulse profile marked in red the Stark pulse chirping smoothly through resonance near $t=50$. The corresponding trajectories on the Bloch sphere are independent of the position $z$, which suggests that the optimal control fields depicted in Figure \ref{fig:zdep}a guarantee the optimal population transfer for all positions in the range $[z_{min},z_{max}]$. This observation indicates that there exists a margin for Stark-chirp rates within which the rapid-adiabatic passage can be completed.

\subsection*{Time and space dependent perturbation}

In general, the Stark pulse can be perturbed with spatially inhomogeneous and time varying fields. The response fields are then written as given below
 \begin{equation}
\begin{split}
\tilde{\Delta}(t;z)=\left(1+f(t)\epsilon(z)\right)\Delta(t) \\
\tilde{\Omega}(t)=\Omega(t)
\end{split}
\label{eq:generalpert}
\end{equation}
where $f(t)=A\cos(\frac{wt}{t_f-t_i})$ and $\epsilon(z)=\cos(\frac{kz}{z_{max}-z_{min}})$. $A,k,w$ are positive parameters characterizing the perturbation. Note that the amplitude of the perturbation shall be no larger than the magnitude of the initial detuning $S_0$ ($A/S_0 \leq 1$). The optimal control equations are given in the form
\begin{equation}
\begin{split}
\left(1+f(t)\epsilon(z)\right)l_3(t)+p_0\frac{\partial r(t,z)}{\partial \Delta}=0 \\
l_1(t)+p_0\frac{\partial r(t;z)}{\partial \Omega}=0
\end{split}
\label{eq:generalopt}
\end{equation}
where $r(t;z)=\int_{z_{min}}^{z_{max}}dz\left(\tilde{\Delta}^2(t;z)+\tilde{\Omega}^2(t;z)\right)$ and $p_0=-1/2$.

The parameters of the perturbation affect the optimal control fields, whereas the control fields dictate the form of the optimal trajectories, which in turn determine the values of the chosen efficiency map, which schematically can be written as $A,w,k \mapsto \Delta^*,\Omega^* \mapsto \mathbf{R},\mathbf{p} \mapsto P(T)$.

\begin{figure}
\includegraphics[width=8.5cm]{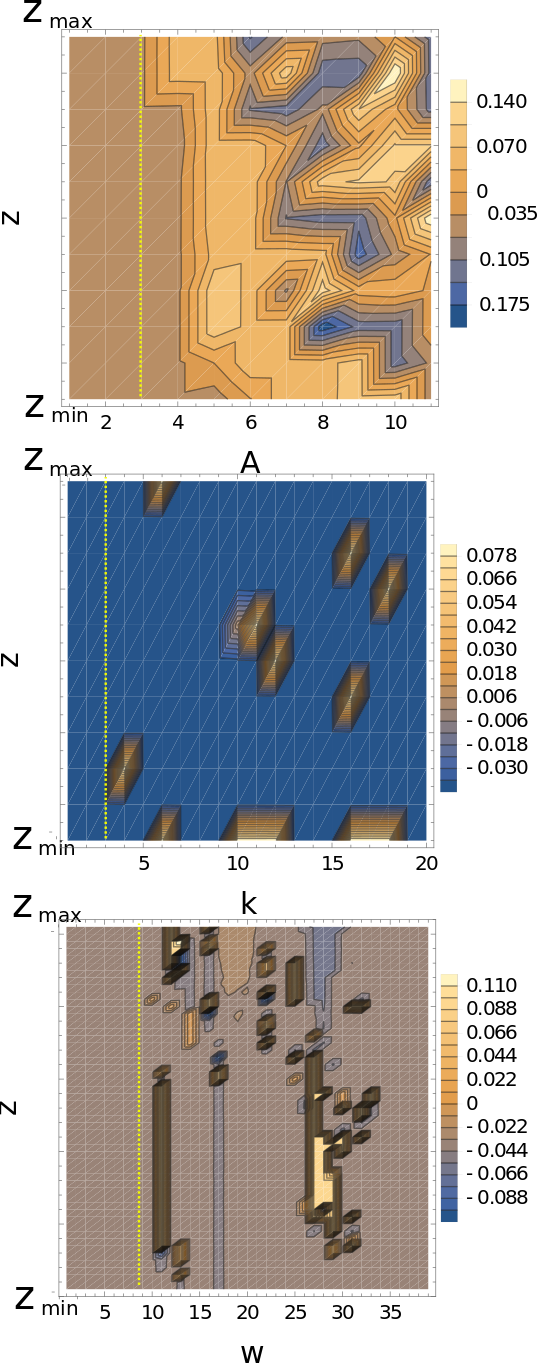}
\caption{Stability maps for the control field $\Delta^*(\tau)$ as a function of: a) z-position and A-amplitude of the perturbing field; b)  z-position and k-wavevector of the perturbing field; c)  z-position and w-frequency of the perturbing field; Dotted yellow lines denote critical values for respective parameters. }
\label{fig:stabilitymap}
\end{figure}

The optimal control pulses depend indirectly on position $z$, which generally means that for every $z$ one should prepare a separate pulse shape. This could be inconvenient from the experimental point of view. Fortunately, there are regions in the $z$-space for which the optimized control fields are independent of $z$. Such region will be called the stability region for optimal controls. 

Infeasibility of certain $z$-coordinate regions is visible in Figure \ref{fig:zdep}, where time profiles for $R_3(t;z)$ are plotted for several values of $z$. Positions, which cover the maxima of the perturbing field (shown in Figure \ref{fig:zdep}d, near $z=0.6Z$) yield slightly different time profiles from the rest of positions. This suggests that the perturbation, at the time of chirping through resonance, is too strong. Thus it is impossible for the optimal population transfer to be $z$-globally complete with a single pair of control fields. This exemplifies limitations on the optimization of control fields, in the environment of inhomogeneous electric fields.

Figure \ref{fig:stabilitymap} depicts the values of the Stark pulse at $t=0.4(t_f-t_i)$ as a function of position $z$ and parameters $A,k,w$ of the perturbing field. This measure indicates if the pulse shape changed for different values of the $z,k,w$ or $A$ parameters. Whenever a pair of the control pulses can be found unchanged for the whole range of positions $z\in[z_{min},z_{max}]$, the SCRAP sequence can be generated with a single optimal pulse, i.e. is experimentally feasible.

The acceptable stability regions in the $A,k,w$ parameters' space form a hexahedron, dimensions of which are marked with yellow dashed lines in Figure \ref{fig:stabilitymap}. All perturbations generated with parameters from inside this acceptance hexahedron result in controllable, $z$-independent optimal shapes for the Stark pulse and the pump pulse.

\subsection*{Adiabatic control}

\begin{figure}
\includegraphics[width=8.5cm]{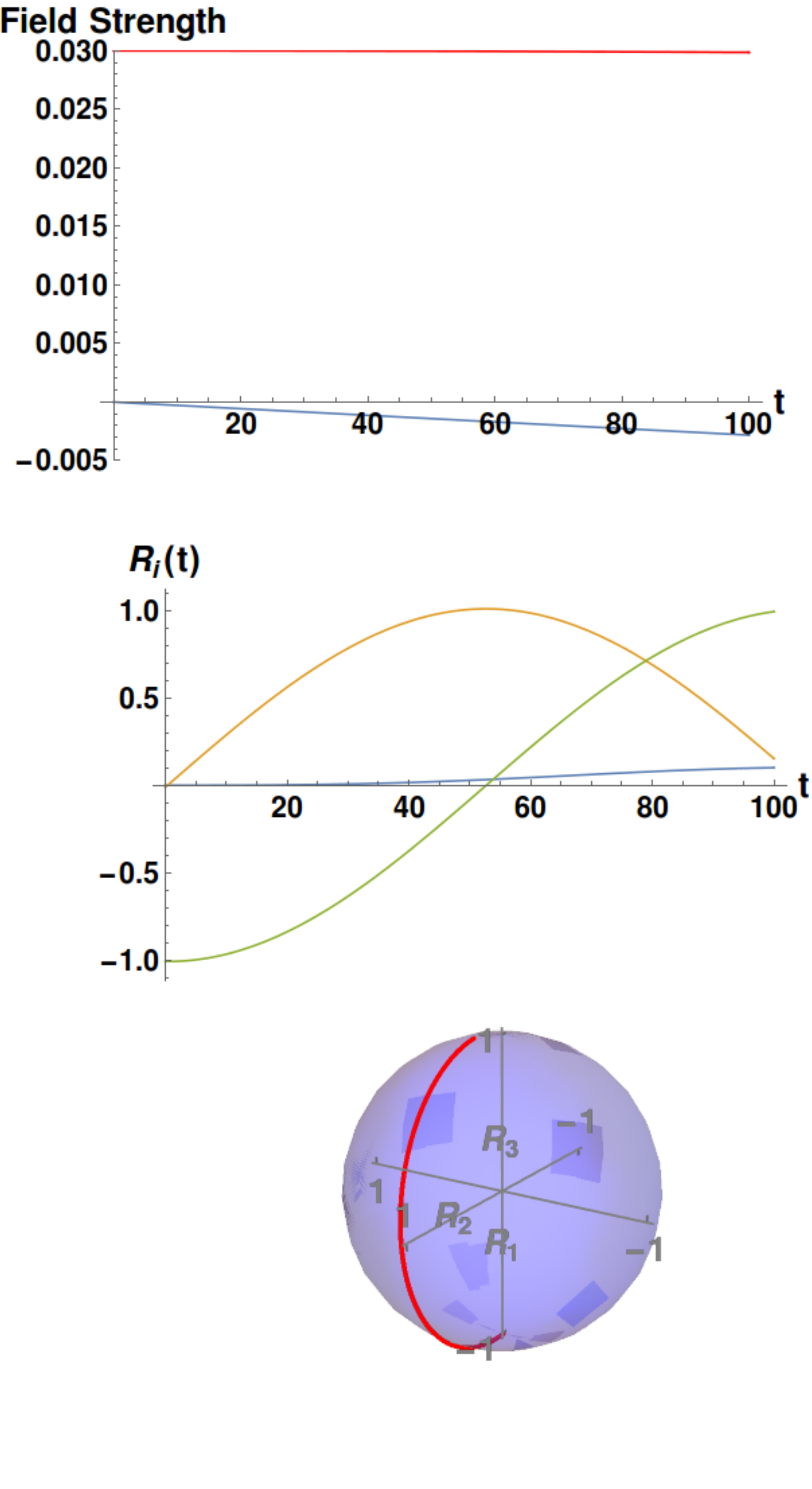}
\caption{a) Optimal control fields calculated for a mixed adiabatic-energy minimum cost functional $r(\vec{\alpha}(t),t)=\dot{\Delta}\Omega-\dot{\Omega}\Delta + \Delta^2+\Omega^2$ used in the PMP procedure; b) Time profiles for the components of the Bloch vector; c) Bloch vector trajectory on the Bloch sphere.}
\label{fig:adopt}
\end{figure}

Results presented in Figure \ref{fig:NABC} and \ref{fig:gpmp} suggest that the adiabatic condition $AD(t)\ll 1$ is not always satisfied for the PMP's optimal control fields.
Thus, the adiabaticity function could be used as a part of the cost functional: $r(\vec{\alpha}(t),t)=AD(t)$. 
The equations for optimal control fields become then a system of eight coupled non-linear differential equations, which are difficult to solve.  A simpler cost functional $r(\vec{\alpha}(t),t)=\dot{\Delta}\Omega-\dot{\Omega}\Delta + \Delta^2+\Omega^2$, which can be viewed as a mixture of the total energy of the pulses and the adiabaticity of the state vector evolution, yields the dynamics shown in Figure \ref{fig:adopt}.

The modified cost functional provides a balance between the energy minimization condition $\min[\Delta^2+\Omega^2]$ and the quasi-adiabatic condition $\min[\dot{\Delta}\Omega-\dot{\Omega}\Delta]$. The former implies the weakest coupling between the adiabatic surfaces, hence pushes the solutions to pass as near as possible to the conical intersection, whereas the latter condition prevents form too rapid oscillations in $\Delta$ and $\Omega$ and pushes trajectories away from the conical intersection. 

Figure \ref{fig:adopt} shows that the quasi-adiabatic condition indeed imposes a smooth time evolution of the components of the Bloch vector, and the associated trajectory on the Bloch sphere is nearly aligned with the adiabatic trajectory. The control fields, which generate such dynamics are presented in Figure \ref{fig:adopt}a. The constant time profile for the pump pulse and a time-linear chirp resemble the Landau-Zener-Stuckelberg model \cite{Vitanov1996,Vasilev2007}.

One of the obvious advantages of using the optimal control theory in SCRAP is its ability to almost instantly provide pulse shapes, which satisfy given conditions and restrictions. It should be noted however that, at times, the generated optimal control fields could not be easily prepared in the laboratory. For this reason, solutions obtained from maximization of the PMP's pseudo-Hamiltonian should be always critically assessed.

\section{The geometric phase in SCRAP}

\begin{figure}
\includegraphics[width=8.5cm]{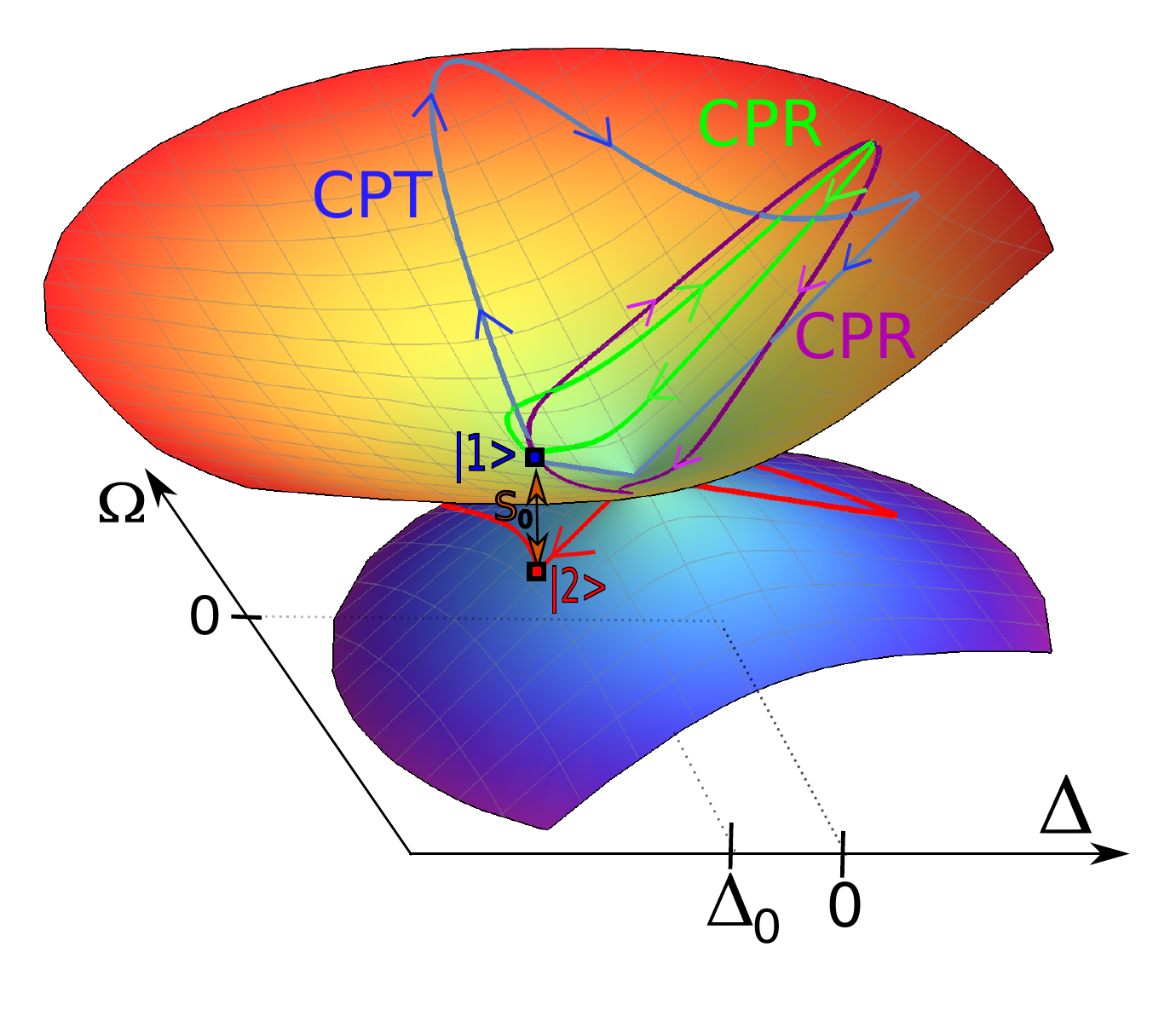}
\caption{Rapid-adiabatic passage dynamics on adiabatic surfaces. The green and purple trajectories correspond to two realisations of the complete population return. The blue path on the upper surface and the red path on the lower surface constitute the complete population transfer ($|1\rangle \rightarrow |2\rangle$) in a typical SCRAP. Initial detuning between the diabatic states is marked with $\Delta_0$ and the initial energetic separation between the diabatic states is denoted as $S_0$. }
\label{fig:CI}
\end{figure}

When an adiabatic evolution of the system completes a closed path in the parameters' space of the Hamiltonian and encircles a conical intersection of adiabatic energy surfaces (cf. Figure \ref{fig:CI}), then a geometric phase is acquired by the adiabatic wavefunction \cite{Berry1984,Mead1992}, i.e. if $H(0)=H(T)$ for some time $T$, then $\psi(T)=e^{i\gamma}\psi(0)$. This additional phase is independent of the dynamical phase of the wavefunction \cite{Berry1984}.

In the realm of rapid-adiabatic passage, there exist scenarios in which the parameters of the Hamiltonian (cf. \ref{eq:RWA}) follow a periodic path. This occurs, for instance, in the complete population return (CPR).

The purple path in Figure \ref{fig:CI} shows such situation, when the CPR path encircles the conical intersection between the two adiabatic energy surfaces and completes a closed path. A topologically different situation occurs for a CPR avoids the conical-intersection, as marked by green line in Figure \ref{fig:CI}. In the former case, upon the completion of the population return, the adiabatic wavefunciton should gain a phase factor, which is independent of path taken by the pump and the stark pulse during the evolution (provided that it is adiabatic). 

Because it has been an increasingly investigated topic in the context of ultra-cold science \cite{Min2014,Zygelman2015,Kendrick2015} (hence the manipulation of qubits too \cite{Zhang2017}), the role of the geometric phase in the adiabatic dynamics deserves a discussion in the context of SCRAP.

The geometric phase can be calculated as the integral over a closed path followed by the vector potential $\vec{A}(t)= \langle \tilde{\phi}_{\pm}|\vec{\nabla}_{\vec{\alpha}} \tilde{\phi}_{\pm}\rangle $:

\begin{equation}
\gamma =i\oint_C \langle \tilde{\phi}_{\pm}|\vec{\nabla}_{\vec{\alpha}} \tilde{\phi}_{\pm}\rangle\cdot d\vec{\alpha}
\label{eq:gp}
\end{equation}
where $\tilde{\phi}_{\pm}=e^{i\theta (t)}\phi_{\pm}$ is the complex adiabatic function unitarily connected \cite{Mead1992} to the original adiabatic functions defined earlier in eq. \ref{eq:AD-DB}. In the above equation $\vec{\alpha}=(\Delta,\Omega)^T$, and $C$ is any closed path in the space of $(\Delta(t),\Omega(t))$ parameters, which starts at $t_i$ and ends at $t_f$. The vector potential $\vec{A}(\vec{\alpha})$ can be calculated directly 

\begin{equation}
\vec{A}(t)=\frac{i}{2}\frac{1}{\Delta^2(t)+\Omega^2(t)}\left( {\begin{array}{c}
   -\Omega(t) \\
    \Delta(t) \\
  \end{array} } \right)
\label{eq:gp1}
\end{equation}
The expression for the geometric phase can be further written as:
\begin{equation}
\gamma =-\frac{i}{2}\int_{t_i}^{t_f} \frac{\Delta(t)\dot{\Omega}(t)-\Omega(t)\dot{\Delta}(t)}{\Delta^2(t)+\Omega^2(t)}dt
\label{eq:gp2}
\end{equation}

For the optimal control pulses which minimize the total energy of the pulse, hence satisfy $\Delta^2+\Omega^2=l_1(t)^2+l_3(t)^2=l^2$, the geometric phase can be calculated directly as an integral over the adiabaticity function:

\begin{equation}
\gamma =-\frac{i}{|l|}\int_{t_i}^{t_f} AD(t) dt
\label{eq:gp3}
\end{equation}
which closely connects the adiabaticity measure of the evolution $AD(t)$ with the geometric phase. Whenever the trajectory followed by the control fields $\Delta(t), \Omega(t)$ encircles the CI located at  $\Delta=0, \Omega=0$, as shown in Figure \ref{fig:CI}, the geometric phase is equal to $\pi$. Otherwise it vanishes. Indeed, any Gaussian time profiles for $\Delta(t)$ and $\Omega(t)$ which encircle the CI gives $\gamma = \pi$. Similarly, for circular paths, which minimize the total energy of the pulses combined, i.e. when the control pulses are parametrized by: $\Delta(t)=r\sin(t), \Omega(t)=r\cos(t)$, it is straightforward to show that the expression in eq. \ref{eq:gp1} is $\pi$, independent of the radius $r$.

\begin{figure}
\includegraphics[width=8.5cm]{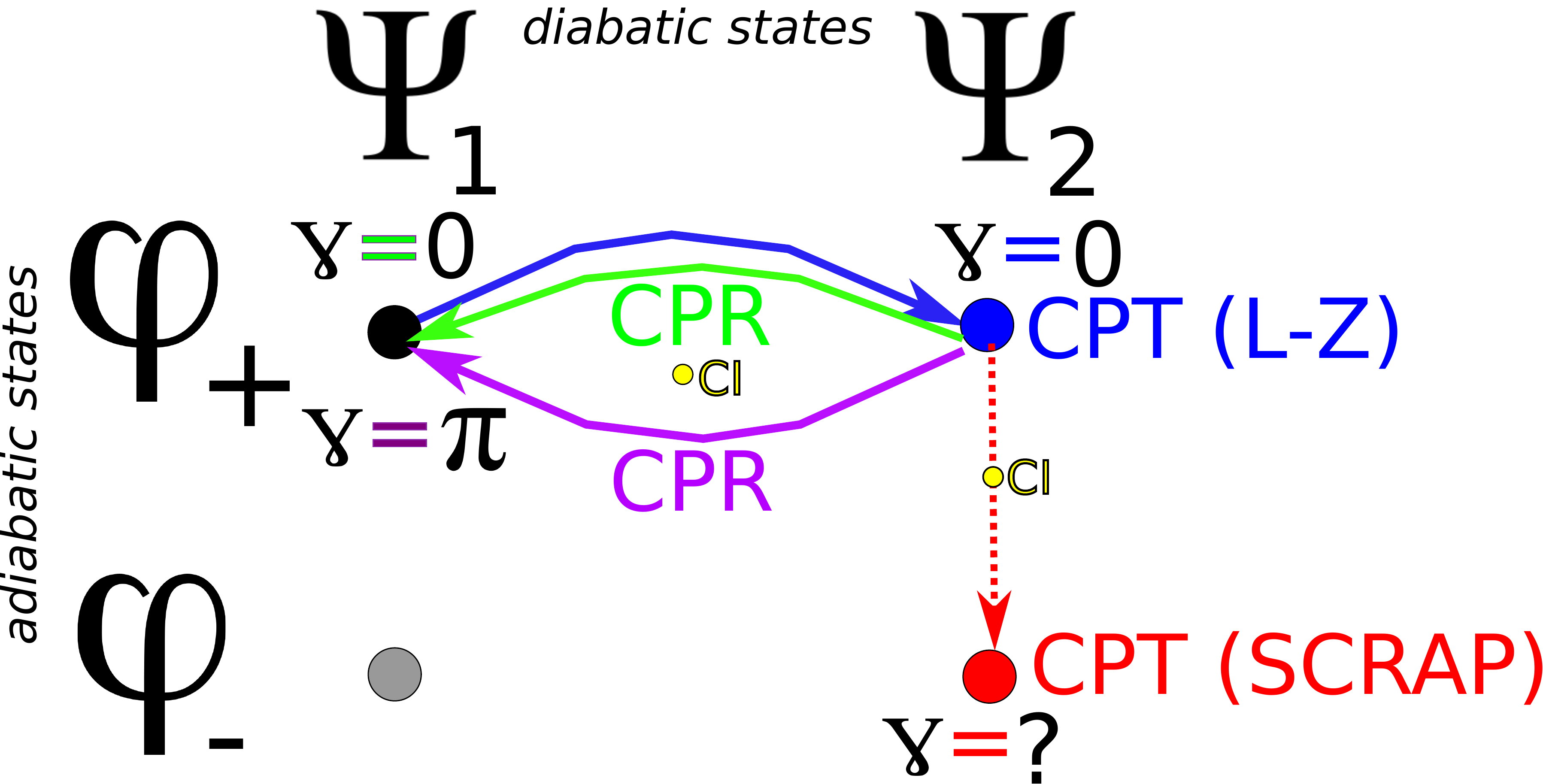}
\caption{A schematic graph presenting the topology of the rapid-adiabatic passage in the adiabatic picture. The green CPR trajectory avoids the conical intersection (CI), and yields zero geometric phase. The purple trajectory goes around the CI and corresponds to the non-zero geometric phase ($\gamma=\pi$). The blue path leads from the initial diabatic state $\psi_1 \equiv |1\rangle$ to $\psi_2 \equiv |2\rangle$, which is qualitatively identical to realisation of the Landau-Zener transition (L-Z). The diabatic evolution with the pump pulse switched off causes passage through the CI and completes the SCRAP sequence.}
\label{fig:evolution}
\end{figure}

A remaining question is whether the geometric phase persists during the complete population transfer $|1\rangle \rightarrow |2\rangle$, as indicated schematically in Figure \ref{fig:evolution}. The time evolution in such a situation is adiabatic at the first resonance, until the second \textit{mute resonance}, when the evolution becomes diabatic, which means that the path goes through or very near the conical intersection, i.e. there is a transition between the adiabatic surfaces. An open question is how does the phase of the target state after the SCRAP sequence relates to the phase of the initial state. Apart from the dynamical phase, the geometric phase should appear due to periodic trajectory followed by the parameters of the Hamiltonian. In such a case, the final state should not be $|2\rangle$, but rather $e^{i\gamma}|2\rangle$. Similarly, after the complete population return the initial and the final state should differ by the geometric phase. Complications can appear in the region of diabatic evolution. The adiabatic theorem \cite{Mead1992} and the related theorem about the geometric phase \cite{Berry1984} assumes adiabatic evolution of the state vector. In the CPR case it is possible to satisfy the adiabatic condition, as shown in Figure \ref{fig:CI}. The purple CPR path on the adiabatic energy surface goes around the conical intersection and returns to its starting point. Such situation precisely satisfies the assumptions of the theorem of the existence of the geometric phase. Does the breakdown of the adiabatic approximation near the mute resonance during CPT interfere with the value of the geometric phase? Is the geometric phase an artifact of the adiabatic approximation? These questions are designated to future studies.

Control fields, which follow a circular path given by the equation $\Delta^2+\Omega^2=l^2=\textit{const}$ minimize the total energy of the pulse, but also explore the region of negative Rabi frequency. In principle, realisation of such a scenario could be possible by chirping not only the intensity of the Stark pulse, but also its polarization angle from 0 degrees to 180 degrees. 
Other effects contributing to the value of the off-diagonal elements in the Hamiltonian can shift the coupling terms into the negative values. For example, vibronic coupling in molecules, which in the linear diabatic model is proportional to, for instance the bond-length, could possibly cause such shifts and induce a non-zero geometric phase. The dynamical consequences carried by the existence of a non-zero geometric phase during CPR is a subject for future studies.

The advantage of dynamical couplings induced by $\Delta(t)$ and $\Omega(t)$ lies in vast freedom of tailoring the pulses, hence probing practically any point or path on the adiabatic surfaces. Conclusions obtained with this highly controllable technique (SCRAP) can be used in nuclear motion theory of multiple electronic states to explain the relation between the geometric phase and the vibronic coupling in molecules.

\section*{Summary and outlook}
The present study focused on the problem of finding the optimal control electric fields for the Stark-chirp rapid-adiabatic passage technique.

In the first instance, an adiabatic model for SCRAP was considered, in which the direct search for the global maximum of the efficiency of the population transfer was performed over the landscape of the Gaussian-shaped Stark pulse parameters. The global maximum of the efficiency map was found to correspond to the dynamics for which the state vector evolves non-adiabatically.  
The breakdown of the adiabatic approximation however does not necessarily implies an incomplete population transfer between the initial and the target state. Results of the numerically exact dynamics calculations were compared with results of calculations in the adiabatic approximation. The calculated exact trajectories of the Bloch vector on the Bloch sphere suggest that a nearly complete population transfer occurs even though the evolution is not fully adiabatic. The residual oscillations of the components of the Bloch vector during this evolution were quantified with the introduced non-adiabatic correction to the adiabatic Bloch vector. This measure informs on the level of adiabaticity of the process, and can be utilized as a diagnostic tool in other studies on the rapid-adiabatic passage.

The second situation, in which no \textit{a priori} assumptions about the functional form of the control fields were made, required the application of the Pontryagin maximum principle.
Two types of cost functionals were tested: the total energy of the pulses and a combination of the total energy of the pulses and the time-derivative of the mixing angle $\Theta$ between the diabatic states. For the former functional, both spatially inhomogeneous and time-dependent perturbing electric fields were considered. For the given perturbing fields, the PMP provided optimal control fields, which ensured robustness SCRAP within a limited range of the perturbation strengths and frequencies. 

The optimization procedures presented in this work apply wherever time and/or position dependent electric inhomogeneities affect the rapid-adiabatic passage.
For instance, spatially inhomogeneous electric fields which affect the chirp rate of the Stark pulse can be present in cavities used for production of ultra-cold molecules, multi-qubit ensembles or chambers for reactive scattering. 

Finally, it was shown that the adiabatic wavefunction acquires a non-zero geometric phase during a complete population return. The existence of a non-zero geometric phase in the complete population transfer, during which the dynamics passes through the diabatic regime, remains still an open question. 

Future studies should concern analysis of the role of the geometric phase and the vibronic coupling in molecules during the rapid-adiabatic passage. The subtle effect of the geometric phase may become significant when SCRAP is employed to manipulate quantum states near a light-induced conical intersection between electronic energy surfaces in molecules.

\section*{Appendix A}
Analytic solutions to equations (\ref{eq:GGcond}) in section \ref{sec:gauss}  are given below. The equations are written as
\begin{equation}
\begin{split}
 \frac{\partial P_{2}(T;t_s,\sigma_s)}{\partial t_s}=\frac{\Omega^2}{4\sigma_s^2}\frac{\Delta}{\left(\Omega^2+\Delta^2\right)^{\frac{3}{2}}}\left(T-t_s\right)=0 \\ 
  \frac{\partial P_{2}(T;t_s,\sigma_s)}{\partial \sigma_s}=-\frac{\Omega}{2\Delta}\frac{1}{1+\left(\frac{\Omega}{\Delta}\right)^2}\left(\frac{(t-t_s)^2}{\sigma_s^3}-\frac{1}{\sigma_s}\right)=0
\label{eq:A1}
\end{split}
\end{equation}
Calculation of the partial derivatives  with respect to $t_s$ and $\sigma_s$ give a set of two algebraic equations, which define two sets of curves. Differentiation of the efficiency function $P_2$ with respect to $t_s$ results in a pair of solution curves $t_s^{\pm}(\sigma_s)$ which are given as:
\begin{equation}
t_s^{\pm}(\sigma_s)=T\pm \sqrt{2\sigma_s^2\log\left(\frac{\sqrt{2\pi}S_0\sigma_s}{\Delta_0}\right)}
\label{eq:A3}
\end{equation}
Differentiation of the efficiency function with respect to pulse width gives two further curves:
\begin{equation}
\sigma_s^{\pm}(t_s)=\pm(T-t_s)
\label{eq:A4}
\end{equation}
Intersections of the two sets of the obtained curves generate a set of three optimal points in the $(t_s,\sigma_s)$ space, depicted in Figure \ref{fig:GGpulses} and discussed in section \ref{sec:gauss}.

\bibliographystyle{apsrev4-1}
\bibliography{scrap}

\begin{thebibliography}{64}%
\makeatletter
\providecommand \@ifxundefined [1]{%
 \@ifx{#1\undefined}
}%
\providecommand \@ifnum [1]{%
 \ifnum #1\expandafter \@firstoftwo
 \else \expandafter \@secondoftwo
 \fi
}%
\providecommand \@ifx [1]{%
 \ifx #1\expandafter \@firstoftwo
 \else \expandafter \@secondoftwo
 \fi
}%
\providecommand \natexlab [1]{#1}%
\providecommand \enquote  [1]{``#1''}%
\providecommand \bibnamefont  [1]{#1}%
\providecommand \bibfnamefont [1]{#1}%
\providecommand \citenamefont [1]{#1}%
\providecommand \href@noop [0]{\@secondoftwo}%
\providecommand \href [0]{\begingroup \@sanitize@url \@href}%
\providecommand \@href[1]{\@@startlink{#1}\@@href}%
\providecommand \@@href[1]{\endgroup#1\@@endlink}%
\providecommand \@sanitize@url [0]{\catcode `\\12\catcode `\$12\catcode
  `\&12\catcode `\#12\catcode `\^12\catcode `\_12\catcode `\%12\relax}%
\providecommand \@@startlink[1]{}%
\providecommand \@@endlink[0]{}%
\providecommand \url  [0]{\begingroup\@sanitize@url \@url }%
\providecommand \@url [1]{\endgroup\@href {#1}{\urlprefix }}%
\providecommand \urlprefix  [0]{URL }%
\providecommand \Eprint [0]{\href }%
\providecommand \doibase [0]{http://dx.doi.org/}%
\providecommand \selectlanguage [0]{\@gobble}%
\providecommand \bibinfo  [0]{\@secondoftwo}%
\providecommand \bibfield  [0]{\@secondoftwo}%
\providecommand \translation [1]{[#1]}%
\providecommand \BibitemOpen [0]{}%
\providecommand \bibitemStop [0]{}%
\providecommand \bibitemNoStop [0]{.\EOS\space}%
\providecommand \EOS [0]{\spacefactor3000\relax}%
\providecommand \BibitemShut  [1]{\csname bibitem#1\endcsname}%
\let\auto@bib@innerbib\@empty
%</preamble>
\bibitem [{\citenamefont {Nie}\ \emph {et~al.}(2010)\citenamefont {Nie},
  \citenamefont {Huang}, \citenamefont {Shi},\ and\ \citenamefont
  {Wei}}]{Nie2010}%
  \BibitemOpen
  \bibfield  {author} {\bibinfo {author} {\bibfnamefont {W.}~\bibnamefont
  {Nie}}, \bibinfo {author} {\bibfnamefont {J.~S.}\ \bibnamefont {Huang}},
  \bibinfo {author} {\bibfnamefont {X.}~\bibnamefont {Shi}}, \ and\ \bibinfo
  {author} {\bibfnamefont {L.~F.}\ \bibnamefont {Wei}},\ }\href {\doibase
  10.1103/physreva.82.032319} {\bibfield  {journal} {\bibinfo  {journal} {Phys.
  Rev. A}\ }\textbf {\bibinfo {volume} {82}} (\bibinfo {year} {2010}),\
  10.1103/physreva.82.032319}\BibitemShut {NoStop}%
\bibitem [{\citenamefont {Chen}\ and\ \citenamefont {Wei}(2015)}]{Chen2015}%
  \BibitemOpen
  \bibfield  {author} {\bibinfo {author} {\bibfnamefont {J.}~\bibnamefont
  {Chen}}\ and\ \bibinfo {author} {\bibfnamefont {L.}~\bibnamefont {Wei}},\
  }\href {\doibase 10.1016/j.physleta.2015.05.035} {\bibfield  {journal}
  {\bibinfo  {journal} {Physics Letters A}\ }\textbf {\bibinfo {volume}
  {379}},\ \bibinfo {pages} {2549} (\bibinfo {year} {2015})}\BibitemShut
  {NoStop}%
\bibitem [{\citenamefont {Rowland}\ and\ \citenamefont
  {Jones}(2012)}]{10.2307/41739851}%
  \BibitemOpen
  \bibfield  {author} {\bibinfo {author} {\bibfnamefont {B.}~\bibnamefont
  {Rowland}}\ and\ \bibinfo {author} {\bibfnamefont {J.~A.}\ \bibnamefont
  {Jones}},\ }\href {http://www.jstor.org/stable/41739851} {\bibfield
  {journal} {\bibinfo  {journal} {Philosophical Transactions: Mathematical,
  Physical and Engineering Sciences}\ }\textbf {\bibinfo {volume} {370}},\
  \bibinfo {pages} {4636} (\bibinfo {year} {2012})}\BibitemShut {NoStop}%
\bibitem [{\citenamefont {Shi}\ \emph {et~al.}(2014)\citenamefont {Shi},
  \citenamefont {Oh},\ and\ \citenamefont {Wei}}]{Shi2014}%
  \BibitemOpen
  \bibfield  {author} {\bibinfo {author} {\bibfnamefont {X.}~\bibnamefont
  {Shi}}, \bibinfo {author} {\bibfnamefont {C.}~\bibnamefont {Oh}}, \ and\
  \bibinfo {author} {\bibfnamefont {L.-F.}\ \bibnamefont {Wei}},\ }\href
  {\doibase 10.1088/0253-6102/61/2/15} {\bibfield  {journal} {\bibinfo
  {journal} {Communications in Theoretical Physics}\ }\textbf {\bibinfo
  {volume} {61}},\ \bibinfo {pages} {235} (\bibinfo {year} {2014})}\BibitemShut
  {NoStop}%
\bibitem [{\citenamefont {Dolde}\ \emph {et~al.}(2014)\citenamefont {Dolde},
  \citenamefont {Bergholm}, \citenamefont {Wang}, \citenamefont {Jakobi},
  \citenamefont {Naydenov}, \citenamefont {Pezzagna}, \citenamefont {Meijer},
  \citenamefont {Jelezko}, \citenamefont {Neumann}, \citenamefont
  {Schulte-Herbr\"{u}ggen}, \citenamefont {Biamonte},\ and\ \citenamefont
  {Wrachtrup}}]{Dolde2014}%
  \BibitemOpen
  \bibfield  {author} {\bibinfo {author} {\bibfnamefont {F.}~\bibnamefont
  {Dolde}}, \bibinfo {author} {\bibfnamefont {V.}~\bibnamefont {Bergholm}},
  \bibinfo {author} {\bibfnamefont {Y.}~\bibnamefont {Wang}}, \bibinfo {author}
  {\bibfnamefont {I.}~\bibnamefont {Jakobi}}, \bibinfo {author} {\bibfnamefont
  {B.}~\bibnamefont {Naydenov}}, \bibinfo {author} {\bibfnamefont
  {S.}~\bibnamefont {Pezzagna}}, \bibinfo {author} {\bibfnamefont
  {J.}~\bibnamefont {Meijer}}, \bibinfo {author} {\bibfnamefont
  {F.}~\bibnamefont {Jelezko}}, \bibinfo {author} {\bibfnamefont
  {P.}~\bibnamefont {Neumann}}, \bibinfo {author} {\bibfnamefont
  {T.}~\bibnamefont {Schulte-Herbr\"{u}ggen}}, \bibinfo {author} {\bibfnamefont
  {J.}~\bibnamefont {Biamonte}}, \ and\ \bibinfo {author} {\bibfnamefont
  {J.}~\bibnamefont {Wrachtrup}},\ }\href {\doibase 10.1038/ncomms4371}
  {\bibfield  {journal} {\bibinfo  {journal} {Nature Communications}\ }\textbf
  {\bibinfo {volume} {5}} (\bibinfo {year} {2014}),\
  10.1038/ncomms4371}\BibitemShut {NoStop}%
\bibitem [{\citenamefont {Huang}\ and\ \citenamefont {Goan}(2014)}]{Huang2014}%
  \BibitemOpen
  \bibfield  {author} {\bibinfo {author} {\bibfnamefont {S.-Y.}\ \bibnamefont
  {Huang}}\ and\ \bibinfo {author} {\bibfnamefont {H.-S.}\ \bibnamefont
  {Goan}},\ }\href {\doibase 10.1103/physreva.90.012318} {\bibfield  {journal}
  {\bibinfo  {journal} {Physical Review A}\ }\textbf {\bibinfo {volume} {90}}
  (\bibinfo {year} {2014}),\ 10.1103/physreva.90.012318}\BibitemShut {NoStop}%
\bibitem [{\citenamefont {Kobzar}\ \emph {et~al.}(2008)\citenamefont {Kobzar},
  \citenamefont {Skinner}, \citenamefont {Khaneja}, \citenamefont {Glaser},\
  and\ \citenamefont {Luy}}]{Kobzar2008}%
  \BibitemOpen
  \bibfield  {author} {\bibinfo {author} {\bibfnamefont {K.}~\bibnamefont
  {Kobzar}}, \bibinfo {author} {\bibfnamefont {T.~E.}\ \bibnamefont {Skinner}},
  \bibinfo {author} {\bibfnamefont {N.}~\bibnamefont {Khaneja}}, \bibinfo
  {author} {\bibfnamefont {S.~J.}\ \bibnamefont {Glaser}}, \ and\ \bibinfo
  {author} {\bibfnamefont {B.}~\bibnamefont {Luy}},\ }\href {\doibase
  10.1016/j.jmr.2008.05.023} {\bibfield  {journal} {\bibinfo  {journal}
  {Journal of Magnetic Resonance}\ }\textbf {\bibinfo {volume} {194}},\
  \bibinfo {pages} {58} (\bibinfo {year} {2008})}\BibitemShut {NoStop}%
\bibitem [{\citenamefont {Kobzar}\ \emph {et~al.}(2012)\citenamefont {Kobzar},
  \citenamefont {Ehni}, \citenamefont {Skinner}, \citenamefont {Glaser},\ and\
  \citenamefont {Luy}}]{Kobzar2012}%
  \BibitemOpen
  \bibfield  {author} {\bibinfo {author} {\bibfnamefont {K.}~\bibnamefont
  {Kobzar}}, \bibinfo {author} {\bibfnamefont {S.}~\bibnamefont {Ehni}},
  \bibinfo {author} {\bibfnamefont {T.~E.}\ \bibnamefont {Skinner}}, \bibinfo
  {author} {\bibfnamefont {S.~J.}\ \bibnamefont {Glaser}}, \ and\ \bibinfo
  {author} {\bibfnamefont {B.}~\bibnamefont {Luy}},\ }\href {\doibase
  10.1016/j.jmr.2012.09.013} {\bibfield  {journal} {\bibinfo  {journal}
  {Journal of Magnetic Resonance}\ }\textbf {\bibinfo {volume} {225}},\
  \bibinfo {pages} {142} (\bibinfo {year} {2012})}\BibitemShut {NoStop}%
\bibitem [{\citenamefont {Reeth}\ \emph {et~al.}(2017)\citenamefont {Reeth},
  \citenamefont {Ratiney}, \citenamefont {Lapert}, \citenamefont {Glaser},\
  and\ \citenamefont {Sugny}}]{VanReeth2017}%
  \BibitemOpen
  \bibfield  {author} {\bibinfo {author} {\bibfnamefont {E.~V.}\ \bibnamefont
  {Reeth}}, \bibinfo {author} {\bibfnamefont {H.}~\bibnamefont {Ratiney}},
  \bibinfo {author} {\bibfnamefont {M.}~\bibnamefont {Lapert}}, \bibinfo
  {author} {\bibfnamefont {S.~J.}\ \bibnamefont {Glaser}}, \ and\ \bibinfo
  {author} {\bibfnamefont {D.}~\bibnamefont {Sugny}},\ }\href {\doibase
  10.1186/s40736-017-0034-3} {\bibfield  {journal} {\bibinfo  {journal}
  {Pacific Journal of Mathematics for Industry}\ }\textbf {\bibinfo {volume}
  {9}} (\bibinfo {year} {2017}),\ 10.1186/s40736-017-0034-3}\BibitemShut
  {NoStop}%
\bibitem [{\citenamefont {Hornung}\ \emph {et~al.}(2000)\citenamefont
  {Hornung}, \citenamefont {Meier},\ and\ \citenamefont
  {Motzkus}}]{Hornung2000}%
  \BibitemOpen
  \bibfield  {author} {\bibinfo {author} {\bibfnamefont {T.}~\bibnamefont
  {Hornung}}, \bibinfo {author} {\bibfnamefont {R.}~\bibnamefont {Meier}}, \
  and\ \bibinfo {author} {\bibfnamefont {M.}~\bibnamefont {Motzkus}},\ }\href
  {\doibase 10.1016/s0009-2614(00)00810-1} {\bibfield  {journal} {\bibinfo
  {journal} {Chemical Physics Letters}\ }\textbf {\bibinfo {volume} {326}},\
  \bibinfo {pages} {445} (\bibinfo {year} {2000})}\BibitemShut {NoStop}%
\bibitem [{\citenamefont {Ren}\ \emph {et~al.}(2006)\citenamefont {Ren},
  \citenamefont {Balint-Kurti}, \citenamefont {Manby}, \citenamefont
  {Artamonov}, \citenamefont {Ho},\ and\ \citenamefont {Rabitz}}]{Ren2006}%
  \BibitemOpen
  \bibfield  {author} {\bibinfo {author} {\bibfnamefont {Q.}~\bibnamefont
  {Ren}}, \bibinfo {author} {\bibfnamefont {G.~G.}\ \bibnamefont
  {Balint-Kurti}}, \bibinfo {author} {\bibfnamefont {F.~R.}\ \bibnamefont
  {Manby}}, \bibinfo {author} {\bibfnamefont {M.}~\bibnamefont {Artamonov}},
  \bibinfo {author} {\bibfnamefont {T.-S.}\ \bibnamefont {Ho}}, \ and\ \bibinfo
  {author} {\bibfnamefont {H.}~\bibnamefont {Rabitz}},\ }\href {\doibase
  10.1063/1.2141616} {\bibfield  {journal} {\bibinfo  {journal} {The Journal of
  Chemical Physics}\ }\textbf {\bibinfo {volume} {124}},\ \bibinfo {pages}
  {014111} (\bibinfo {year} {2006})}\BibitemShut {NoStop}%
\bibitem [{\citenamefont {Nuernberger}\ \emph {et~al.}(2007)\citenamefont
  {Nuernberger}, \citenamefont {Vogt}, \citenamefont {Brixner},\ and\
  \citenamefont {Gerber}}]{Nuernberger2007}%
  \BibitemOpen
  \bibfield  {author} {\bibinfo {author} {\bibfnamefont {P.}~\bibnamefont
  {Nuernberger}}, \bibinfo {author} {\bibfnamefont {G.}~\bibnamefont {Vogt}},
  \bibinfo {author} {\bibfnamefont {T.}~\bibnamefont {Brixner}}, \ and\
  \bibinfo {author} {\bibfnamefont {G.}~\bibnamefont {Gerber}},\ }\href
  {\doibase 10.1039/b618760a} {\bibfield  {journal} {\bibinfo  {journal}
  {Physical Chemistry Chemical Physics}\ }\textbf {\bibinfo {volume} {9}},\
  \bibinfo {pages} {2470} (\bibinfo {year} {2007})}\BibitemShut {NoStop}%
\bibitem [{\citenamefont {Brif}\ \emph {et~al.}(2011)\citenamefont {Brif},
  \citenamefont {Chakrabarti},\ and\ \citenamefont {Rabitz}}]{Brif2011}%
  \BibitemOpen
  \bibfield  {author} {\bibinfo {author} {\bibfnamefont {C.}~\bibnamefont
  {Brif}}, \bibinfo {author} {\bibfnamefont {R.}~\bibnamefont {Chakrabarti}}, \
  and\ \bibinfo {author} {\bibfnamefont {H.}~\bibnamefont {Rabitz}},\ }in\
  \href {\doibase 10.1002/9781118158715.ch1} {\emph {\bibinfo {booktitle}
  {Advances in Chemical Physics}}}\ (\bibinfo  {publisher} {John Wiley {\&}
  Sons, Inc.},\ \bibinfo {year} {2011})\ pp.\ \bibinfo {pages}
  {1--76}\BibitemShut {NoStop}%
\bibitem [{\citenamefont {Wells}\ \emph {et~al.}(2013)\citenamefont {Wells},
  \citenamefont {Rallis}, \citenamefont {Zohrabi}, \citenamefont {Siemering},
  \citenamefont {Jochim}, \citenamefont {Andrews}, \citenamefont {Ablikim},
  \citenamefont {Gaire}, \citenamefont {De}, \citenamefont {Carnes},
  \citenamefont {Bergues}, \citenamefont {de~Vivie-Riedle}, \citenamefont
  {Kling},\ and\ \citenamefont {Ben-Itzhak}}]{Wells2013}%
  \BibitemOpen
  \bibfield  {author} {\bibinfo {author} {\bibfnamefont {E.}~\bibnamefont
  {Wells}}, \bibinfo {author} {\bibfnamefont {C.}~\bibnamefont {Rallis}},
  \bibinfo {author} {\bibfnamefont {M.}~\bibnamefont {Zohrabi}}, \bibinfo
  {author} {\bibfnamefont {R.}~\bibnamefont {Siemering}}, \bibinfo {author}
  {\bibfnamefont {B.}~\bibnamefont {Jochim}}, \bibinfo {author} {\bibfnamefont
  {P.}~\bibnamefont {Andrews}}, \bibinfo {author} {\bibfnamefont
  {U.}~\bibnamefont {Ablikim}}, \bibinfo {author} {\bibfnamefont
  {B.}~\bibnamefont {Gaire}}, \bibinfo {author} {\bibfnamefont
  {S.}~\bibnamefont {De}}, \bibinfo {author} {\bibfnamefont {K.}~\bibnamefont
  {Carnes}}, \bibinfo {author} {\bibfnamefont {B.}~\bibnamefont {Bergues}},
  \bibinfo {author} {\bibfnamefont {R.}~\bibnamefont {de~Vivie-Riedle}},
  \bibinfo {author} {\bibfnamefont {M.}~\bibnamefont {Kling}}, \ and\ \bibinfo
  {author} {\bibfnamefont {I.}~\bibnamefont {Ben-Itzhak}},\ }\href {\doibase
  10.1038/ncomms3895} {\bibfield  {journal} {\bibinfo  {journal} {Nature
  Communications}\ }\textbf {\bibinfo {volume} {4}} (\bibinfo {year} {2013}),\
  10.1038/ncomms3895}\BibitemShut {NoStop}%
\bibitem [{\citenamefont {Parker}\ \emph {et~al.}(2014)\citenamefont {Parker},
  \citenamefont {Smeu}, \citenamefont {Franco}, \citenamefont {Ratner},\ and\
  \citenamefont {Seideman}}]{Parker2014}%
  \BibitemOpen
  \bibfield  {author} {\bibinfo {author} {\bibfnamefont {S.~M.}\ \bibnamefont
  {Parker}}, \bibinfo {author} {\bibfnamefont {M.}~\bibnamefont {Smeu}},
  \bibinfo {author} {\bibfnamefont {I.}~\bibnamefont {Franco}}, \bibinfo
  {author} {\bibfnamefont {M.~A.}\ \bibnamefont {Ratner}}, \ and\ \bibinfo
  {author} {\bibfnamefont {T.}~\bibnamefont {Seideman}},\ }\href {\doibase
  10.1021/nl501629c} {\bibfield  {journal} {\bibinfo  {journal} {Nano Letters}\
  }\textbf {\bibinfo {volume} {14}},\ \bibinfo {pages} {4587} (\bibinfo {year}
  {2014})}\BibitemShut {NoStop}%
\bibitem [{\citenamefont {Strasfeld}\ \emph {et~al.}(2007)\citenamefont
  {Strasfeld}, \citenamefont {Shim},\ and\ \citenamefont
  {Zanni}}]{Strasfeld2007}%
  \BibitemOpen
  \bibfield  {author} {\bibinfo {author} {\bibfnamefont {D.~B.}\ \bibnamefont
  {Strasfeld}}, \bibinfo {author} {\bibfnamefont {S.-H.}\ \bibnamefont {Shim}},
  \ and\ \bibinfo {author} {\bibfnamefont {M.~T.}\ \bibnamefont {Zanni}},\
  }\href {\doibase 10.1103/physrevlett.99.038102} {\bibfield  {journal}
  {\bibinfo  {journal} {Physical Review Letters}\ }\textbf {\bibinfo {volume}
  {99}} (\bibinfo {year} {2007}),\ 10.1103/physrevlett.99.038102}\BibitemShut
  {NoStop}%
\bibitem [{\citenamefont {Sugny}\ \emph {et~al.}(2009)\citenamefont {Sugny},
  \citenamefont {Bomble}, \citenamefont {Ribeyre}, \citenamefont {Dulieu},\
  and\ \citenamefont {Desouter-Lecomte}}]{Sugny2009}%
  \BibitemOpen
  \bibfield  {author} {\bibinfo {author} {\bibfnamefont {D.}~\bibnamefont
  {Sugny}}, \bibinfo {author} {\bibfnamefont {L.}~\bibnamefont {Bomble}},
  \bibinfo {author} {\bibfnamefont {T.}~\bibnamefont {Ribeyre}}, \bibinfo
  {author} {\bibfnamefont {O.}~\bibnamefont {Dulieu}}, \ and\ \bibinfo {author}
  {\bibfnamefont {M.}~\bibnamefont {Desouter-Lecomte}},\ }\href {\doibase
  10.1103/physreva.80.042325} {\bibfield  {journal} {\bibinfo  {journal}
  {Physical Review A}\ }\textbf {\bibinfo {volume} {80}} (\bibinfo {year}
  {2009}),\ 10.1103/physreva.80.042325}\BibitemShut {NoStop}%
\bibitem [{\citenamefont {Sch\"{o}nfeldt}\ \emph {et~al.}(2009)\citenamefont
  {Sch\"{o}nfeldt}, \citenamefont {Twamley},\ and\ \citenamefont
  {Rebi{\'{c}}}}]{Schnfeldt2009}%
  \BibitemOpen
  \bibfield  {author} {\bibinfo {author} {\bibfnamefont {J.-H.}\ \bibnamefont
  {Sch\"{o}nfeldt}}, \bibinfo {author} {\bibfnamefont {J.}~\bibnamefont
  {Twamley}}, \ and\ \bibinfo {author} {\bibfnamefont {S.}~\bibnamefont
  {Rebi{\'{c}}}},\ }\href {\doibase 10.1103/physreva.80.043401} {\bibfield
  {journal} {\bibinfo  {journal} {Physical Review A}\ }\textbf {\bibinfo
  {volume} {80}} (\bibinfo {year} {2009}),\
  10.1103/physreva.80.043401}\BibitemShut {NoStop}%
\bibitem [{\citenamefont {Dong}\ \emph {et~al.}(2013)\citenamefont {Dong},
  \citenamefont {Mukherjee},\ and\ \citenamefont {Zare}}]{Dong2013}%
  \BibitemOpen
  \bibfield  {author} {\bibinfo {author} {\bibfnamefont {W.}~\bibnamefont
  {Dong}}, \bibinfo {author} {\bibfnamefont {N.}~\bibnamefont {Mukherjee}}, \
  and\ \bibinfo {author} {\bibfnamefont {R.~N.}\ \bibnamefont {Zare}},\ }\href
  {\doibase 10.1063/1.4818526} {\bibfield  {journal} {\bibinfo  {journal} {The
  Journal of Chemical Physics}\ }\textbf {\bibinfo {volume} {139}},\ \bibinfo
  {pages} {074204} (\bibinfo {year} {2013})}\BibitemShut {NoStop}%
\bibitem [{\citenamefont {Brif}\ \emph {et~al.}(2014)\citenamefont {Brif},
  \citenamefont {Grace}, \citenamefont {Sarovar},\ and\ \citenamefont
  {Young}}]{Brif2014}%
  \BibitemOpen
  \bibfield  {author} {\bibinfo {author} {\bibfnamefont {C.}~\bibnamefont
  {Brif}}, \bibinfo {author} {\bibfnamefont {M.~D.}\ \bibnamefont {Grace}},
  \bibinfo {author} {\bibfnamefont {M.}~\bibnamefont {Sarovar}}, \ and\
  \bibinfo {author} {\bibfnamefont {K.~C.}\ \bibnamefont {Young}},\ }\href
  {\doibase 10.1088/1367-2630/16/6/065013} {\bibfield  {journal} {\bibinfo
  {journal} {New Journal of Physics}\ }\textbf {\bibinfo {volume} {16}},\
  \bibinfo {pages} {065013} (\bibinfo {year} {2014})}\BibitemShut {NoStop}%
\bibitem [{\citenamefont {Pelzer}\ \emph {et~al.}(2007)\citenamefont {Pelzer},
  \citenamefont {Ramakrishna},\ and\ \citenamefont {Seideman}}]{Pelzer2007}%
  \BibitemOpen
  \bibfield  {author} {\bibinfo {author} {\bibfnamefont {A.}~\bibnamefont
  {Pelzer}}, \bibinfo {author} {\bibfnamefont {S.}~\bibnamefont {Ramakrishna}},
  \ and\ \bibinfo {author} {\bibfnamefont {T.}~\bibnamefont {Seideman}},\
  }\href {\doibase 10.1063/1.2408423} {\bibfield  {journal} {\bibinfo
  {journal} {The Journal of Chemical Physics}\ }\textbf {\bibinfo {volume}
  {126}},\ \bibinfo {pages} {034503} (\bibinfo {year} {2007})}\BibitemShut
  {NoStop}%
\bibitem [{\citenamefont {Suzuki}\ \emph {et~al.}(2008)\citenamefont {Suzuki},
  \citenamefont {Sugawara}, \citenamefont {Minemoto},\ and\ \citenamefont
  {Sakai}}]{Suzuki2008}%
  \BibitemOpen
  \bibfield  {author} {\bibinfo {author} {\bibfnamefont {T.}~\bibnamefont
  {Suzuki}}, \bibinfo {author} {\bibfnamefont {Y.}~\bibnamefont {Sugawara}},
  \bibinfo {author} {\bibfnamefont {S.}~\bibnamefont {Minemoto}}, \ and\
  \bibinfo {author} {\bibfnamefont {H.}~\bibnamefont {Sakai}},\ }\href
  {\doibase 10.1103/physrevlett.100.033603} {\bibfield  {journal} {\bibinfo
  {journal} {Physical Review Letters}\ }\textbf {\bibinfo {volume} {100}}
  (\bibinfo {year} {2008}),\ 10.1103/physrevlett.100.033603}\BibitemShut
  {NoStop}%
\bibitem [{\citenamefont {Artamonov}\ and\ \citenamefont
  {Seideman}(2010)}]{Artamonov2010}%
  \BibitemOpen
  \bibfield  {author} {\bibinfo {author} {\bibfnamefont {M.}~\bibnamefont
  {Artamonov}}\ and\ \bibinfo {author} {\bibfnamefont {T.}~\bibnamefont
  {Seideman}},\ }\href {\doibase 10.1103/physreva.82.023413} {\bibfield
  {journal} {\bibinfo  {journal} {Physical Review A}\ }\textbf {\bibinfo
  {volume} {82}} (\bibinfo {year} {2010}),\
  10.1103/physreva.82.023413}\BibitemShut {NoStop}%
\bibitem [{\citenamefont {Lin}\ \emph {et~al.}(2011)\citenamefont {Lin},
  \citenamefont {Zhang}, \citenamefont {Cai},\ and\ \citenamefont
  {Chen}}]{Lin2011}%
  \BibitemOpen
  \bibfield  {author} {\bibinfo {author} {\bibfnamefont {Y.}~\bibnamefont
  {Lin}}, \bibinfo {author} {\bibfnamefont {Z.}~\bibnamefont {Zhang}}, \bibinfo
  {author} {\bibfnamefont {S.}~\bibnamefont {Cai}}, \ and\ \bibinfo {author}
  {\bibfnamefont {Z.}~\bibnamefont {Chen}},\ }\href {\doibase
  10.1021/ja1113479} {\bibfield  {journal} {\bibinfo  {journal} {Journal of the
  American Chemical Society}\ }\textbf {\bibinfo {volume} {133}},\ \bibinfo
  {pages} {7632} (\bibinfo {year} {2011})}\BibitemShut {NoStop}%
\bibitem [{\citenamefont {Kurosaki}\ \emph {et~al.}(2009)\citenamefont
  {Kurosaki}, \citenamefont {Artamonov}, \citenamefont {Ho},\ and\
  \citenamefont {Rabitz}}]{Kurosaki2009}%
  \BibitemOpen
  \bibfield  {author} {\bibinfo {author} {\bibfnamefont {Y.}~\bibnamefont
  {Kurosaki}}, \bibinfo {author} {\bibfnamefont {M.}~\bibnamefont {Artamonov}},
  \bibinfo {author} {\bibfnamefont {T.-S.}\ \bibnamefont {Ho}}, \ and\ \bibinfo
  {author} {\bibfnamefont {H.}~\bibnamefont {Rabitz}},\ }\href {\doibase
  10.1063/1.3185565} {\bibfield  {journal} {\bibinfo  {journal} {The Journal of
  Chemical Physics}\ }\textbf {\bibinfo {volume} {131}},\ \bibinfo {pages}
  {044306} (\bibinfo {year} {2009})}\BibitemShut {NoStop}%
\bibitem [{\citenamefont {Krieger}\ \emph {et~al.}(2011)\citenamefont
  {Krieger}, \citenamefont {Castro},\ and\ \citenamefont
  {Gross}}]{Krieger2011}%
  \BibitemOpen
  \bibfield  {author} {\bibinfo {author} {\bibfnamefont {K.}~\bibnamefont
  {Krieger}}, \bibinfo {author} {\bibfnamefont {A.}~\bibnamefont {Castro}}, \
  and\ \bibinfo {author} {\bibfnamefont {E.}~\bibnamefont {Gross}},\ }\href
  {\doibase 10.1016/j.chemphys.2011.04.014} {\bibfield  {journal} {\bibinfo
  {journal} {Chemical Physics}\ }\textbf {\bibinfo {volume} {391}},\ \bibinfo
  {pages} {50} (\bibinfo {year} {2011})}\BibitemShut {NoStop}%
\bibitem [{\citenamefont {Pan}\ \emph {et~al.}(2017)\citenamefont {Pan},
  \citenamefont {Mondal}, \citenamefont {Yang},\ and\ \citenamefont
  {Liu}}]{Pan2017}%
  \BibitemOpen
  \bibfield  {author} {\bibinfo {author} {\bibfnamefont {H.}~\bibnamefont
  {Pan}}, \bibinfo {author} {\bibfnamefont {S.}~\bibnamefont {Mondal}},
  \bibinfo {author} {\bibfnamefont {C.-H.}\ \bibnamefont {Yang}}, \ and\
  \bibinfo {author} {\bibfnamefont {K.}~\bibnamefont {Liu}},\ }\href {\doibase
  10.1063/1.4982615} {\bibfield  {journal} {\bibinfo  {journal} {The Journal of
  Chemical Physics}\ }\textbf {\bibinfo {volume} {147}},\ \bibinfo {pages}
  {013928} (\bibinfo {year} {2017})}\BibitemShut {NoStop}%
\bibitem [{\citenamefont {Jurdjevic}(1996)}]{jurdjevic}%
  \BibitemOpen
  \bibfield  {author} {\bibinfo {author} {\bibfnamefont {V.}~\bibnamefont
  {Jurdjevic}}\ }(\bibinfo  {publisher} {Cambridge University Press,
  Cambridge},\ \bibinfo {year} {1996})\BibitemShut {NoStop}%
\bibitem [{\citenamefont {Bonnard}\ and\ \citenamefont
  {Sugny}(2012)}]{Bonnard2012}%
  \BibitemOpen
  \bibfield  {author} {\bibinfo {author} {\bibfnamefont {B.}~\bibnamefont
  {Bonnard}}\ and\ \bibinfo {author} {\bibfnamefont {D.}~\bibnamefont
  {Sugny}},\ }in\ \href@noop {} {\emph {\bibinfo {booktitle} {AIMS Appl.
  Math.}}}\ (\bibinfo  {publisher} {American Institute of Mathematical
  Sciences},\ \bibinfo {year} {2012})\ p.\ \bibinfo {pages} {Vol.
  5}\BibitemShut {NoStop}%
\bibitem [{\citenamefont {Werschnik}\ and\ \citenamefont
  {Gross}(2007)}]{Werschnik2007}%
  \BibitemOpen
  \bibfield  {author} {\bibinfo {author} {\bibfnamefont {J.}~\bibnamefont
  {Werschnik}}\ and\ \bibinfo {author} {\bibfnamefont {E.~K.~U.}\ \bibnamefont
  {Gross}},\ }\href {\doibase 10.1088/0953-4075/40/18/r01} {\bibfield
  {journal} {\bibinfo  {journal} {Journal of Physics B: Atomic, Molecular and
  Optical Physics}\ }\textbf {\bibinfo {volume} {40}},\ \bibinfo {pages} {R175}
  (\bibinfo {year} {2007})}\BibitemShut {NoStop}%
\bibitem [{\citenamefont {Reich}\ \emph {et~al.}(2012)\citenamefont {Reich},
  \citenamefont {Ndong},\ and\ \citenamefont {Koch}}]{Reich2012}%
  \BibitemOpen
  \bibfield  {author} {\bibinfo {author} {\bibfnamefont {D.~M.}\ \bibnamefont
  {Reich}}, \bibinfo {author} {\bibfnamefont {M.}~\bibnamefont {Ndong}}, \ and\
  \bibinfo {author} {\bibfnamefont {C.~P.}\ \bibnamefont {Koch}},\ }\href
  {\doibase 10.1063/1.3691827} {\bibfield  {journal} {\bibinfo  {journal} {The
  Journal of Chemical Physics}\ }\textbf {\bibinfo {volume} {136}},\ \bibinfo
  {pages} {104103} (\bibinfo {year} {2012})}\BibitemShut {NoStop}%
\bibitem [{\citenamefont {Riviello}\ \emph {et~al.}(2017)\citenamefont
  {Riviello}, \citenamefont {Wu}, \citenamefont {Sun},\ and\ \citenamefont
  {Rabitz}}]{Riviello2017}%
  \BibitemOpen
  \bibfield  {author} {\bibinfo {author} {\bibfnamefont {G.}~\bibnamefont
  {Riviello}}, \bibinfo {author} {\bibfnamefont {R.-B.}\ \bibnamefont {Wu}},
  \bibinfo {author} {\bibfnamefont {Q.}~\bibnamefont {Sun}}, \ and\ \bibinfo
  {author} {\bibfnamefont {H.}~\bibnamefont {Rabitz}},\ }\href {\doibase
  10.1103/physreva.95.063418} {\bibfield  {journal} {\bibinfo  {journal}
  {Physical Review A}\ }\textbf {\bibinfo {volume} {95}} (\bibinfo {year}
  {2017}),\ 10.1103/physreva.95.063418}\BibitemShut {NoStop}%
\bibitem [{\citenamefont {Boscain}\ \emph {et~al.}(2002)\citenamefont
  {Boscain}, \citenamefont {Charlot}, \citenamefont {Gauthier}, \citenamefont
  {Guerin},\ and\ \citenamefont {Jauslin}}]{Boscain2002}%
  \BibitemOpen
  \bibfield  {author} {\bibinfo {author} {\bibfnamefont {U.}~\bibnamefont
  {Boscain}}, \bibinfo {author} {\bibfnamefont {G.}~\bibnamefont {Charlot}},
  \bibinfo {author} {\bibfnamefont {J.-P.}\ \bibnamefont {Gauthier}}, \bibinfo
  {author} {\bibfnamefont {S.}~\bibnamefont {Guerin}}, \ and\ \bibinfo {author}
  {\bibfnamefont {H.-R.}\ \bibnamefont {Jauslin}},\ }\href {\doibase
  10.1063/1.1465516} {\bibfield  {journal} {\bibinfo  {journal} {Journal of
  Mathematical Physics}\ }\textbf {\bibinfo {volume} {43}},\ \bibinfo {pages}
  {2107} (\bibinfo {year} {2002})}\BibitemShut {NoStop}%
\bibitem [{\citenamefont {Ass{\'{e}}mat}\ and\ \citenamefont
  {Sugny}(2012)}]{Assmat2012}%
  \BibitemOpen
  \bibfield  {author} {\bibinfo {author} {\bibfnamefont {E.}~\bibnamefont
  {Ass{\'{e}}mat}}\ and\ \bibinfo {author} {\bibfnamefont {D.}~\bibnamefont
  {Sugny}},\ }\href {\doibase 10.1103/physreva.86.023406} {\bibfield  {journal}
  {\bibinfo  {journal} {Physical Review A}\ }\textbf {\bibinfo {volume} {86}}
  (\bibinfo {year} {2012}),\ 10.1103/physreva.86.023406}\BibitemShut {NoStop}%
\bibitem [{\citenamefont {Damme}\ \emph {et~al.}(2017)\citenamefont {Damme},
  \citenamefont {Ansel}, \citenamefont {Glaser},\ and\ \citenamefont
  {Sugny}}]{VanDamme2017}%
  \BibitemOpen
  \bibfield  {author} {\bibinfo {author} {\bibfnamefont {L.~V.}\ \bibnamefont
  {Damme}}, \bibinfo {author} {\bibfnamefont {Q.}~\bibnamefont {Ansel}},
  \bibinfo {author} {\bibfnamefont {S.~J.}\ \bibnamefont {Glaser}}, \ and\
  \bibinfo {author} {\bibfnamefont {D.}~\bibnamefont {Sugny}},\ }\href
  {\doibase 10.1103/physreva.95.063403} {\bibfield  {journal} {\bibinfo
  {journal} {Physical Review A}\ }\textbf {\bibinfo {volume} {95}} (\bibinfo
  {year} {2017}),\ 10.1103/physreva.95.063403}\BibitemShut {NoStop}%
\bibitem [{\citenamefont {Rickes}\ \emph {et~al.}(2000)\citenamefont {Rickes},
  \citenamefont {Yatsenko}, \citenamefont {Steuerwald}, \citenamefont
  {Halfmann}, \citenamefont {Shore}, \citenamefont {Vitanov},\ and\
  \citenamefont {Bergmann}}]{Rickes2000}%
  \BibitemOpen
  \bibfield  {author} {\bibinfo {author} {\bibfnamefont {T.}~\bibnamefont
  {Rickes}}, \bibinfo {author} {\bibfnamefont {L.~P.}\ \bibnamefont
  {Yatsenko}}, \bibinfo {author} {\bibfnamefont {S.}~\bibnamefont
  {Steuerwald}}, \bibinfo {author} {\bibfnamefont {T.}~\bibnamefont
  {Halfmann}}, \bibinfo {author} {\bibfnamefont {B.~W.}\ \bibnamefont {Shore}},
  \bibinfo {author} {\bibfnamefont {N.~V.}\ \bibnamefont {Vitanov}}, \ and\
  \bibinfo {author} {\bibfnamefont {K.}~\bibnamefont {Bergmann}},\ }\href
  {\doibase 10.1063/1.481829} {\bibfield  {journal} {\bibinfo  {journal} {The
  Journal of Chemical Physics}\ }\textbf {\bibinfo {volume} {113}},\ \bibinfo
  {pages} {534} (\bibinfo {year} {2000})}\BibitemShut {NoStop}%
\bibitem [{\citenamefont {Yatsenko}\ \emph {et~al.}(1999)\citenamefont
  {Yatsenko}, \citenamefont {Shore}, \citenamefont {Halfmann}, \citenamefont
  {Bergmann},\ and\ \citenamefont {Vardi}}]{Yatsenko1999}%
  \BibitemOpen
  \bibfield  {author} {\bibinfo {author} {\bibfnamefont {L.~P.}\ \bibnamefont
  {Yatsenko}}, \bibinfo {author} {\bibfnamefont {B.~W.}\ \bibnamefont {Shore}},
  \bibinfo {author} {\bibfnamefont {T.}~\bibnamefont {Halfmann}}, \bibinfo
  {author} {\bibfnamefont {K.}~\bibnamefont {Bergmann}}, \ and\ \bibinfo
  {author} {\bibfnamefont {A.}~\bibnamefont {Vardi}},\ }\href {\doibase
  10.1103/physreva.60.r4237} {\bibfield  {journal} {\bibinfo  {journal} {Phys.
  Rev. A}\ }\textbf {\bibinfo {volume} {60}},\ \bibinfo {pages} {R4237}
  (\bibinfo {year} {1999})}\BibitemShut {NoStop}%
\bibitem [{\citenamefont {Oberst}\ \emph {et~al.}(2008)\citenamefont {Oberst},
  \citenamefont {Munch}, \citenamefont {Grigoryan},\ and\ \citenamefont
  {Halfmann}}]{Oberst2008}%
  \BibitemOpen
  \bibfield  {author} {\bibinfo {author} {\bibfnamefont {M.}~\bibnamefont
  {Oberst}}, \bibinfo {author} {\bibfnamefont {H.}~\bibnamefont {Munch}},
  \bibinfo {author} {\bibfnamefont {G.}~\bibnamefont {Grigoryan}}, \ and\
  \bibinfo {author} {\bibfnamefont {T.}~\bibnamefont {Halfmann}},\ }\href
  {\doibase 10.1103/physreva.78.033409} {\bibfield  {journal} {\bibinfo
  {journal} {Phys. Rev. A}\ }\textbf {\bibinfo {volume} {78}} (\bibinfo {year}
  {2008}),\ 10.1103/physreva.78.033409}\BibitemShut {NoStop}%
\bibitem [{\citenamefont {Shore}\ \emph {et~al.}(2009)\citenamefont {Shore},
  \citenamefont {Gromovyy}, \citenamefont {Yatsenko},\ and\ \citenamefont
  {Romanenko}}]{Shore2009}%
  \BibitemOpen
  \bibfield  {author} {\bibinfo {author} {\bibfnamefont {B.~W.}\ \bibnamefont
  {Shore}}, \bibinfo {author} {\bibfnamefont {M.~V.}\ \bibnamefont {Gromovyy}},
  \bibinfo {author} {\bibfnamefont {L.~P.}\ \bibnamefont {Yatsenko}}, \ and\
  \bibinfo {author} {\bibfnamefont {V.~I.}\ \bibnamefont {Romanenko}},\ }\href
  {\doibase 10.1119/1.3231688} {\bibfield  {journal} {\bibinfo  {journal}
  {American Journal of Physics}\ }\textbf {\bibinfo {volume} {77}},\ \bibinfo
  {pages} {1183} (\bibinfo {year} {2009})}\BibitemShut {NoStop}%
\bibitem [{\citenamefont {Vitanov}\ \emph {et~al.}(2017)\citenamefont
  {Vitanov}, \citenamefont {Rangelov}, \citenamefont {Shore},\ and\
  \citenamefont {Bergmann}}]{Vitanov2017}%
  \BibitemOpen
  \bibfield  {author} {\bibinfo {author} {\bibfnamefont {N.~V.}\ \bibnamefont
  {Vitanov}}, \bibinfo {author} {\bibfnamefont {A.~A.}\ \bibnamefont
  {Rangelov}}, \bibinfo {author} {\bibfnamefont {B.~W.}\ \bibnamefont {Shore}},
  \ and\ \bibinfo {author} {\bibfnamefont {K.}~\bibnamefont {Bergmann}},\
  }\href {\doibase 10.1103/revmodphys.89.015006} {\bibfield  {journal}
  {\bibinfo  {journal} {Reviews of Modern Physics}\ }\textbf {\bibinfo {volume}
  {89}} (\bibinfo {year} {2017}),\ 10.1103/revmodphys.89.015006}\BibitemShut
  {NoStop}%
\bibitem [{\citenamefont {Parker}\ \emph {et~al.}(2012)\citenamefont {Parker},
  \citenamefont {Ratner},\ and\ \citenamefont {Seideman}}]{Parker2012}%
  \BibitemOpen
  \bibfield  {author} {\bibinfo {author} {\bibfnamefont {S.~M.}\ \bibnamefont
  {Parker}}, \bibinfo {author} {\bibfnamefont {M.~A.}\ \bibnamefont {Ratner}},
  \ and\ \bibinfo {author} {\bibfnamefont {T.}~\bibnamefont {Seideman}},\
  }\href {\doibase 10.1080/00268976.2012.695808} {\bibfield  {journal}
  {\bibinfo  {journal} {Molecular Physics}\ }\textbf {\bibinfo {volume}
  {110}},\ \bibinfo {pages} {1941} (\bibinfo {year} {2012})}\BibitemShut
  {NoStop}%
\bibitem [{\citenamefont {Chen}\ \emph {et~al.}(2011)\citenamefont {Chen},
  \citenamefont {Torrontegui}, \citenamefont {Stefanatos}, \citenamefont {Li},\
  and\ \citenamefont {Muga}}]{Chen2011}%
  \BibitemOpen
  \bibfield  {author} {\bibinfo {author} {\bibfnamefont {X.}~\bibnamefont
  {Chen}}, \bibinfo {author} {\bibfnamefont {E.}~\bibnamefont {Torrontegui}},
  \bibinfo {author} {\bibfnamefont {D.}~\bibnamefont {Stefanatos}}, \bibinfo
  {author} {\bibfnamefont {J.-S.}\ \bibnamefont {Li}}, \ and\ \bibinfo {author}
  {\bibfnamefont {J.~G.}\ \bibnamefont {Muga}},\ }\href {\doibase
  10.1103/physreva.84.043415} {\bibfield  {journal} {\bibinfo  {journal}
  {Physical Review A}\ }\textbf {\bibinfo {volume} {84}} (\bibinfo {year}
  {2011}),\ 10.1103/physreva.84.043415}\BibitemShut {NoStop}%
\bibitem [{\citenamefont {Gerry}\ and\ \citenamefont
  {Knight}(2004)}]{gerry_knight_2004}%
  \BibitemOpen
  \bibfield  {author} {\bibinfo {author} {\bibfnamefont {C.}~\bibnamefont
  {Gerry}}\ and\ \bibinfo {author} {\bibfnamefont {P.}~\bibnamefont {Knight}},\
  }\href {\doibase 10.1017/CBO9780511791239} {\emph {\bibinfo {title}
  {Introductory Quantum Optics}}}\ (\bibinfo  {publisher} {Cambridge University
  Press},\ \bibinfo {year} {2004})\BibitemShut {NoStop}%
\bibitem [{\citenamefont {Khaneja}\ \emph {et~al.}(2005)\citenamefont
  {Khaneja}, \citenamefont {Reiss}, \citenamefont {Kehlet}, \citenamefont
  {Schulte-Herbr\"{u}ggen},\ and\ \citenamefont {Glaser}}]{Khaneja2005}%
  \BibitemOpen
  \bibfield  {author} {\bibinfo {author} {\bibfnamefont {N.}~\bibnamefont
  {Khaneja}}, \bibinfo {author} {\bibfnamefont {T.}~\bibnamefont {Reiss}},
  \bibinfo {author} {\bibfnamefont {C.}~\bibnamefont {Kehlet}}, \bibinfo
  {author} {\bibfnamefont {T.}~\bibnamefont {Schulte-Herbr\"{u}ggen}}, \ and\
  \bibinfo {author} {\bibfnamefont {S.~J.}\ \bibnamefont {Glaser}},\ }\href
  {\doibase 10.1016/j.jmr.2004.11.004} {\bibfield  {journal} {\bibinfo
  {journal} {Journal of Magnetic Resonance}\ }\textbf {\bibinfo {volume}
  {172}},\ \bibinfo {pages} {296} (\bibinfo {year} {2005})}\BibitemShut
  {NoStop}%
\bibitem [{\citenamefont {Li}\ and\ \citenamefont {Khaneja}(2007)}]{Li2007}%
  \BibitemOpen
  \bibfield  {author} {\bibinfo {author} {\bibfnamefont {J.-S.}\ \bibnamefont
  {Li}}\ and\ \bibinfo {author} {\bibfnamefont {N.}~\bibnamefont {Khaneja}},\
  }\href {\doibase 10.3182/20070822-3-za-2920.00021} {\bibfield  {journal}
  {\bibinfo  {journal} {{IFAC} Proceedings Volumes}\ }\textbf {\bibinfo
  {volume} {40}},\ \bibinfo {pages} {123} (\bibinfo {year} {2007})}\BibitemShut
  {NoStop}%
\bibitem [{\citenamefont {Altafini}(2007)}]{Altafini2007}%
  \BibitemOpen
  \bibfield  {author} {\bibinfo {author} {\bibfnamefont {C.}~\bibnamefont
  {Altafini}},\ }\href {\doibase 10.1109/tac.2007.908306} {\bibfield  {journal}
  {\bibinfo  {journal} {{IEEE} Transactions on Automatic Control}\ }\textbf
  {\bibinfo {volume} {52}},\ \bibinfo {pages} {2019} (\bibinfo {year}
  {2007})}\BibitemShut {NoStop}%
\bibitem [{\citenamefont {Chen}\ \emph {et~al.}(2014)\citenamefont {Chen},
  \citenamefont {Dong}, \citenamefont {Long}, \citenamefont {Petersen},\ and\
  \citenamefont {Rabitz}}]{Chen2014}%
  \BibitemOpen
  \bibfield  {author} {\bibinfo {author} {\bibfnamefont {C.}~\bibnamefont
  {Chen}}, \bibinfo {author} {\bibfnamefont {D.}~\bibnamefont {Dong}}, \bibinfo
  {author} {\bibfnamefont {R.}~\bibnamefont {Long}}, \bibinfo {author}
  {\bibfnamefont {I.~R.}\ \bibnamefont {Petersen}}, \ and\ \bibinfo {author}
  {\bibfnamefont {H.~A.}\ \bibnamefont {Rabitz}},\ }\href {\doibase
  10.1103/physreva.89.023402} {\bibfield  {journal} {\bibinfo  {journal}
  {Physical Review A}\ }\textbf {\bibinfo {volume} {89}} (\bibinfo {year}
  {2014}),\ 10.1103/physreva.89.023402}\BibitemShut {NoStop}%
\bibitem [{\citenamefont {lu~Song}\ \emph {et~al.}(2016)\citenamefont
  {lu~Song}, \citenamefont {li~Yang}, \citenamefont {qi~Yin}, \citenamefont
  {yong Chen},\ and\ \citenamefont {Feng}}]{Song2016}%
  \BibitemOpen
  \bibfield  {author} {\bibinfo {author} {\bibfnamefont {W.}~\bibnamefont
  {lu~Song}}, \bibinfo {author} {\bibfnamefont {W.}~\bibnamefont {li~Yang}},
  \bibinfo {author} {\bibfnamefont {Z.}~\bibnamefont {qi~Yin}}, \bibinfo
  {author} {\bibfnamefont {C.}~\bibnamefont {yong Chen}}, \ and\ \bibinfo
  {author} {\bibfnamefont {M.}~\bibnamefont {Feng}},\ }\href {\doibase
  10.1038/srep33271} {\bibfield  {journal} {\bibinfo  {journal} {Scientific
  Reports}\ }\textbf {\bibinfo {volume} {6}} (\bibinfo {year} {2016}),\
  10.1038/srep33271}\BibitemShut {NoStop}%
\bibitem [{\citenamefont {Lazarou}\ \emph {et~al.}(2010)\citenamefont
  {Lazarou}, \citenamefont {Keller},\ and\ \citenamefont
  {Garraway}}]{Lazarou2010}%
  \BibitemOpen
  \bibfield  {author} {\bibinfo {author} {\bibfnamefont {C.}~\bibnamefont
  {Lazarou}}, \bibinfo {author} {\bibfnamefont {M.}~\bibnamefont {Keller}}, \
  and\ \bibinfo {author} {\bibfnamefont {B.~M.}\ \bibnamefont {Garraway}},\
  }\href {\doibase 10.1103/physreva.81.013418} {\bibfield  {journal} {\bibinfo
  {journal} {Physical Review A}\ }\textbf {\bibinfo {volume} {81}} (\bibinfo
  {year} {2010}),\ 10.1103/physreva.81.013418}\BibitemShut {NoStop}%
\bibitem [{\citenamefont {Parker}\ \emph {et~al.}(2011)\citenamefont {Parker},
  \citenamefont {Ratner},\ and\ \citenamefont {Seideman}}]{Parker2011}%
  \BibitemOpen
  \bibfield  {author} {\bibinfo {author} {\bibfnamefont {S.~M.}\ \bibnamefont
  {Parker}}, \bibinfo {author} {\bibfnamefont {M.~A.}\ \bibnamefont {Ratner}},
  \ and\ \bibinfo {author} {\bibfnamefont {T.}~\bibnamefont {Seideman}},\
  }\href {\doibase 10.1063/1.3663710} {\bibfield  {journal} {\bibinfo
  {journal} {The Journal of Chemical Physics}\ }\textbf {\bibinfo {volume}
  {135}},\ \bibinfo {pages} {224301} (\bibinfo {year} {2011})}\BibitemShut
  {NoStop}%
\bibitem [{\citenamefont {Bethlem}\ \emph {et~al.}(2002)\citenamefont
  {Bethlem}, \citenamefont {van Roij}, \citenamefont {Jongma},\ and\
  \citenamefont {Meijer}}]{Bethlem2002PRL}%
  \BibitemOpen
  \bibfield  {author} {\bibinfo {author} {\bibfnamefont {H.~L.}\ \bibnamefont
  {Bethlem}}, \bibinfo {author} {\bibfnamefont {A.~J.~A.}\ \bibnamefont {van
  Roij}}, \bibinfo {author} {\bibfnamefont {R.~T.}\ \bibnamefont {Jongma}}, \
  and\ \bibinfo {author} {\bibfnamefont {G.}~\bibnamefont {Meijer}},\ }\href
  {\doibase 10.1103/physrevlett.88.133003} {\bibfield  {journal} {\bibinfo
  {journal} {Phys. Rev. Lett.}\ }\textbf {\bibinfo {volume} {88}} (\bibinfo
  {year} {2002}),\ 10.1103/physrevlett.88.133003}\BibitemShut {NoStop}%
\bibitem [{\citenamefont {Bethlem}\ \emph {et~al.}(2006)\citenamefont
  {Bethlem}, \citenamefont {Tarbutt}, \citenamefont {K\"{u}pper}, \citenamefont
  {Carty}, \citenamefont {Wohlfart}, \citenamefont {Hinds},\ and\ \citenamefont
  {Meijer}}]{Kupper2006}%
  \BibitemOpen
  \bibfield  {author} {\bibinfo {author} {\bibfnamefont {H.~L.}\ \bibnamefont
  {Bethlem}}, \bibinfo {author} {\bibfnamefont {M.~R.}\ \bibnamefont
  {Tarbutt}}, \bibinfo {author} {\bibfnamefont {J.}~\bibnamefont {K\"{u}pper}},
  \bibinfo {author} {\bibfnamefont {D.}~\bibnamefont {Carty}}, \bibinfo
  {author} {\bibfnamefont {K.}~\bibnamefont {Wohlfart}}, \bibinfo {author}
  {\bibfnamefont {E.~A.}\ \bibnamefont {Hinds}}, \ and\ \bibinfo {author}
  {\bibfnamefont {G.}~\bibnamefont {Meijer}},\ }\href
  {http://stacks.iop.org/0953-4075/39/i=16/a=R01} {\bibfield  {journal}
  {\bibinfo  {journal} {Journal of Physics B: Atomic, Molecular and Optical
  Physics}\ }\textbf {\bibinfo {volume} {39}},\ \bibinfo {pages} {R263}
  (\bibinfo {year} {2006})}\BibitemShut {NoStop}%
\bibitem [{\citenamefont {Zeppenfeld}\ \emph {et~al.}(2012)\citenamefont
  {Zeppenfeld}, \citenamefont {Englert}, \citenamefont {Gl\"{o}ckner},
  \citenamefont {Prehn}, \citenamefont {Mielenz}, \citenamefont {Sommer},
  \citenamefont {van Buuren}, \citenamefont {Motsch},\ and\ \citenamefont
  {Rempe}}]{Zeppenfeld2012}%
  \BibitemOpen
  \bibfield  {author} {\bibinfo {author} {\bibfnamefont {M.}~\bibnamefont
  {Zeppenfeld}}, \bibinfo {author} {\bibfnamefont {B.~G.~U.}\ \bibnamefont
  {Englert}}, \bibinfo {author} {\bibfnamefont {R.}~\bibnamefont
  {Gl\"{o}ckner}}, \bibinfo {author} {\bibfnamefont {A.}~\bibnamefont {Prehn}},
  \bibinfo {author} {\bibfnamefont {M.}~\bibnamefont {Mielenz}}, \bibinfo
  {author} {\bibfnamefont {C.}~\bibnamefont {Sommer}}, \bibinfo {author}
  {\bibfnamefont {L.~D.}\ \bibnamefont {van Buuren}}, \bibinfo {author}
  {\bibfnamefont {M.}~\bibnamefont {Motsch}}, \ and\ \bibinfo {author}
  {\bibfnamefont {G.}~\bibnamefont {Rempe}},\ }\href {\doibase
  10.1038/nature11595} {\bibfield  {journal} {\bibinfo  {journal} {Nature}\
  }\textbf {\bibinfo {volume} {491}},\ \bibinfo {pages} {570} (\bibinfo {year}
  {2012})}\BibitemShut {NoStop}%
\bibitem [{\citenamefont {Owens}\ \emph {et~al.}(2017)\citenamefont {Owens},
  \citenamefont {Zak}, \citenamefont {Chubb}, \citenamefont {Yurchenko},
  \citenamefont {Tennyson},\ and\ \citenamefont {Yachmenev}}]{Owens2017}%
  \BibitemOpen
  \bibfield  {author} {\bibinfo {author} {\bibfnamefont {A.}~\bibnamefont
  {Owens}}, \bibinfo {author} {\bibfnamefont {E.~J.}\ \bibnamefont {Zak}},
  \bibinfo {author} {\bibfnamefont {K.~L.}\ \bibnamefont {Chubb}}, \bibinfo
  {author} {\bibfnamefont {S.~N.}\ \bibnamefont {Yurchenko}}, \bibinfo {author}
  {\bibfnamefont {J.}~\bibnamefont {Tennyson}}, \ and\ \bibinfo {author}
  {\bibfnamefont {A.}~\bibnamefont {Yachmenev}},\ }\href {\doibase
  10.1038/srep45068} {\bibfield  {journal} {\bibinfo  {journal} {Scientific
  Reports}\ }\textbf {\bibinfo {volume} {7}},\ \bibinfo {pages} {45068}
  (\bibinfo {year} {2017})}\BibitemShut {NoStop}%
\bibitem [{\citenamefont {Shirley}(1965)}]{Shirley1965}%
  \BibitemOpen
  \bibfield  {author} {\bibinfo {author} {\bibfnamefont {J.~H.}\ \bibnamefont
  {Shirley}},\ }\href {\doibase 10.1103/physrev.138.b979} {\bibfield  {journal}
  {\bibinfo  {journal} {Physical Review}\ }\textbf {\bibinfo {volume} {138}},\
  \bibinfo {pages} {B979} (\bibinfo {year} {1965})}\BibitemShut {NoStop}%
\bibitem [{\citenamefont {Rangelov}\ \emph {et~al.}(2005)\citenamefont
  {Rangelov}, \citenamefont {Vitanov}, \citenamefont {Yatsenko}, \citenamefont
  {Shore}, \citenamefont {Halfmann},\ and\ \citenamefont
  {Bergmann}}]{Rangelov2005}%
  \BibitemOpen
  \bibfield  {author} {\bibinfo {author} {\bibfnamefont {A.~A.}\ \bibnamefont
  {Rangelov}}, \bibinfo {author} {\bibfnamefont {N.~V.}\ \bibnamefont
  {Vitanov}}, \bibinfo {author} {\bibfnamefont {L.~P.}\ \bibnamefont
  {Yatsenko}}, \bibinfo {author} {\bibfnamefont {B.~W.}\ \bibnamefont {Shore}},
  \bibinfo {author} {\bibfnamefont {T.}~\bibnamefont {Halfmann}}, \ and\
  \bibinfo {author} {\bibfnamefont {K.}~\bibnamefont {Bergmann}},\ }\href
  {\doibase 10.1103/physreva.72.053403} {\bibfield  {journal} {\bibinfo
  {journal} {Physical Review A}\ }\textbf {\bibinfo {volume} {72}} (\bibinfo
  {year} {2005}),\ 10.1103/physreva.72.053403}\BibitemShut {NoStop}%
\bibitem [{\citenamefont {Vitanov}\ and\ \citenamefont
  {Garraway}(1996)}]{Vitanov1996}%
  \BibitemOpen
  \bibfield  {author} {\bibinfo {author} {\bibfnamefont {N.~V.}\ \bibnamefont
  {Vitanov}}\ and\ \bibinfo {author} {\bibfnamefont {B.~M.}\ \bibnamefont
  {Garraway}},\ }\href {\doibase 10.1103/physreva.53.4288} {\bibfield
  {journal} {\bibinfo  {journal} {Physical Review A}\ }\textbf {\bibinfo
  {volume} {53}},\ \bibinfo {pages} {4288} (\bibinfo {year}
  {1996})}\BibitemShut {NoStop}%
\bibitem [{\citenamefont {Vasilev}\ \emph {et~al.}(2007)\citenamefont
  {Vasilev}, \citenamefont {Ivanov},\ and\ \citenamefont
  {Vitanov}}]{Vasilev2007}%
  \BibitemOpen
  \bibfield  {author} {\bibinfo {author} {\bibfnamefont {G.~S.}\ \bibnamefont
  {Vasilev}}, \bibinfo {author} {\bibfnamefont {S.~S.}\ \bibnamefont {Ivanov}},
  \ and\ \bibinfo {author} {\bibfnamefont {N.~V.}\ \bibnamefont {Vitanov}},\
  }\href {\doibase 10.1103/physreva.75.013417} {\bibfield  {journal} {\bibinfo
  {journal} {Physical Review A}\ }\textbf {\bibinfo {volume} {75}} (\bibinfo
  {year} {2007}),\ 10.1103/physreva.75.013417}\BibitemShut {NoStop}%
\bibitem [{\citenamefont {Berry}(1984)}]{Berry1984}%
  \BibitemOpen
  \bibfield  {author} {\bibinfo {author} {\bibfnamefont {M.~V.}\ \bibnamefont
  {Berry}},\ }\href {\doibase 10.1098/rspa.1984.0023} {\bibfield  {journal}
  {\bibinfo  {journal} {Proceedings of the Royal Society A: Mathematical,
  Physical and Engineering Sciences}\ }\textbf {\bibinfo {volume} {392}},\
  \bibinfo {pages} {45} (\bibinfo {year} {1984})}\BibitemShut {NoStop}%
\bibitem [{\citenamefont {Mead}(1992)}]{Mead1992}%
  \BibitemOpen
  \bibfield  {author} {\bibinfo {author} {\bibfnamefont {C.~A.}\ \bibnamefont
  {Mead}},\ }\href {\doibase 10.1103/revmodphys.64.51} {\bibfield  {journal}
  {\bibinfo  {journal} {Reviews of Modern Physics}\ }\textbf {\bibinfo {volume}
  {64}},\ \bibinfo {pages} {51} (\bibinfo {year} {1992})}\BibitemShut {NoStop}%
\bibitem [{\citenamefont {Min}\ \emph {et~al.}(2014)\citenamefont {Min},
  \citenamefont {Abedi}, \citenamefont {Kim},\ and\ \citenamefont
  {Gross}}]{Min2014}%
  \BibitemOpen
  \bibfield  {author} {\bibinfo {author} {\bibfnamefont {S.~K.}\ \bibnamefont
  {Min}}, \bibinfo {author} {\bibfnamefont {A.}~\bibnamefont {Abedi}}, \bibinfo
  {author} {\bibfnamefont {K.~S.}\ \bibnamefont {Kim}}, \ and\ \bibinfo
  {author} {\bibfnamefont {E.}~\bibnamefont {Gross}},\ }\href {\doibase
  10.1103/physrevlett.113.263004} {\bibfield  {journal} {\bibinfo  {journal}
  {Physical Review Letters}\ }\textbf {\bibinfo {volume} {113}} (\bibinfo
  {year} {2014}),\ 10.1103/physrevlett.113.263004}\BibitemShut {NoStop}%
\bibitem [{\citenamefont {Zygelman}(2015)}]{Zygelman2015}%
  \BibitemOpen
  \bibfield  {author} {\bibinfo {author} {\bibfnamefont {B.}~\bibnamefont
  {Zygelman}},\ }\href {\doibase 10.1103/physreva.92.043620} {\bibfield
  {journal} {\bibinfo  {journal} {Physical Review A}\ }\textbf {\bibinfo
  {volume} {92}} (\bibinfo {year} {2015}),\
  10.1103/physreva.92.043620}\BibitemShut {NoStop}%
\bibitem [{\citenamefont {Kendrick}\ \emph {et~al.}(2015)\citenamefont
  {Kendrick}, \citenamefont {Hazra},\ and\ \citenamefont
  {Balakrishnan}}]{Kendrick2015}%
  \BibitemOpen
  \bibfield  {author} {\bibinfo {author} {\bibfnamefont {B.~K.}\ \bibnamefont
  {Kendrick}}, \bibinfo {author} {\bibfnamefont {J.}~\bibnamefont {Hazra}}, \
  and\ \bibinfo {author} {\bibfnamefont {N.}~\bibnamefont {Balakrishnan}},\
  }\href {\doibase 10.1038/ncomms8918} {\bibfield  {journal} {\bibinfo
  {journal} {Nature Communications}\ }\textbf {\bibinfo {volume} {6}},\
  \bibinfo {pages} {7918} (\bibinfo {year} {2015})}\BibitemShut {NoStop}%
\bibitem [{\citenamefont {Zhang}\ \emph {et~al.}(2017)\citenamefont {Zhang},
  \citenamefont {Wang}, \citenamefont {Xiang}, \citenamefont {Yao},
  \citenamefont {Wu},\ and\ \citenamefont {Yin}}]{Zhang2017}%
  \BibitemOpen
  \bibfield  {author} {\bibinfo {author} {\bibfnamefont {Z.}~\bibnamefont
  {Zhang}}, \bibinfo {author} {\bibfnamefont {T.}~\bibnamefont {Wang}},
  \bibinfo {author} {\bibfnamefont {L.}~\bibnamefont {Xiang}}, \bibinfo
  {author} {\bibfnamefont {J.}~\bibnamefont {Yao}}, \bibinfo {author}
  {\bibfnamefont {J.}~\bibnamefont {Wu}}, \ and\ \bibinfo {author}
  {\bibfnamefont {Y.}~\bibnamefont {Yin}},\ }\href {\doibase
  10.1103/physreva.95.042345} {\bibfield  {journal} {\bibinfo  {journal}
  {Physical Review A}\ }\textbf {\bibinfo {volume} {95}} (\bibinfo {year}
  {2017}),\ 10.1103/physreva.95.042345}\BibitemShut {NoStop}%
\end{thebibliography}%
\end{document}